\documentclass[iop,apj]{emulateapj}

\begin{document}

\newcommand{\gps}{\ensuremath{g_{\rm P1}}}
\newcommand{\rps}{\ensuremath{r_{\rm P1}}}
\newcommand{\ips}{\ensuremath{i_{\rm P1}}}
\newcommand{\zps}{\ensuremath{z_{\rm P1}}}
\newcommand{\yps}{\ensuremath{y_{\rm P1}}}
\newcommand{\wps}{\ensuremath{w_{\rm P1}}}
\newcommand{\grizy}{\gps\rps\ips\zps\yps}
\newcommand{\griz}{\gps\rps\ips\zps}
\newcommand{\gr}{$\gps - \rps$}
\newcommand{\gi}{$\gps - \ips$}
\newcommand{\gz}{$\gps - \zps$}
\newcommand{\PS}{\protect \hbox {Pan-STARRS1}}
\newcommand{\galex}{\ensuremath{GALEX}}
\newcommand{\af}{PS1-11af}
\newcommand{\jh}{PS1-10jh}
\newcommand{\msun}{$M_{\sun}$}
\newcommand{\kms}{km~s$^{-1}$}
\newcommand{\ebv}{$E(B-V)$}
\newcommand{\tone}{SDSS~TDE1}
\newcommand{\ttwo}{SDSS~TDE2}
\newcommand{\swone}{Sw~1644+57}
\newcommand{\swtwo}{Sw~2058+05}
\newcommand{\tbb}{\ensuremath{T_{\mathrm{BB}}}}
\newcommand{\rbb}{\ensuremath{R_{\mathrm{BB}}}}
\newcommand{\lbb}{\ensuremath{L_{\mathrm{BB}}}}
\newcommand{\lbol}{\ensuremath{L_{\mathrm{bol}}}}
\newcommand{\mbol}{\ensuremath{M_{\mathrm{bol}}}}
\newcommand{\ledd}{\ensuremath{L_{\mathrm{Edd}}}}
\newcommand{\mbh}{\ensuremath{M_{\mathrm{bh}}}}
\newcommand{\mdot}{\ensuremath{\dot{M}_{\mathrm{acc}}}}
\newcommand{\mstar}{\ensuremath{M_{\star}}}

\shorttitle{The TDE PS1-11af}
\shortauthors{Chornock et al.}

\title{The UV-bright, Slowly Declining Transient PS1-11\lowercase{af} as
  a Partial Tidal Disruption Event} 

\author{R. Chornock\altaffilmark{1},
E.~Berger\altaffilmark{1},
S.~Gezari\altaffilmark{2},
B.~A.~Zauderer\altaffilmark{1},
A.~Rest\altaffilmark{3},
L.~Chomiuk\altaffilmark{4},
A.~Kamble\altaffilmark{1},
A.~M.~Soderberg\altaffilmark{1},
I.~Czekala\altaffilmark{1},
J.~Dittmann\altaffilmark{1},
M.~Drout\altaffilmark{1},
R.~J.~Foley\altaffilmark{1,5,6,7},
W.~Fong\altaffilmark{1},
M.~E.~Huber\altaffilmark{8},
R.~P.~Kirshner\altaffilmark{1},
A.~Lawrence\altaffilmark{9},
R.~Lunnan\altaffilmark{1},
G.~H.~Marion\altaffilmark{1,10},
G.~Narayan\altaffilmark{1},
A.~G.~Riess\altaffilmark{3,11},
K.~C.~Roth\altaffilmark{12},
N.~E.~Sanders\altaffilmark{1},
D.~Scolnic\altaffilmark{11},
S.~J.~Smartt\altaffilmark{13},
K.~Smith\altaffilmark{13},
C.~W.~Stubbs\altaffilmark{1,14},
J.~L.~Tonry\altaffilmark{8},
W.~S.~Burgett\altaffilmark{8},
K.~C.~Chambers\altaffilmark{8},
H.~Flewelling\altaffilmark{8},
K.~W.~Hodapp\altaffilmark{8},
N. Kaiser\altaffilmark{8},
E.~A.~Magnier\altaffilmark{8},
D.~C.~Martin\altaffilmark{15},
J.~D.~Neill\altaffilmark{15},
P.~A.~Price\altaffilmark{16},
and R.~Wainscoat\altaffilmark{8}
}

\altaffiltext{1}{Harvard-Smithsonian Center for Astrophysics, 60
  Garden St., Cambridge, MA 02138, USA, \texttt{rchornock@cfa.harvard.edu}}
\altaffiltext{2}{Department of Astronomy, University of Maryland,
  College Park, MD 20742-2421, USA} 
\altaffiltext{3}{Space Telescope Science Institute, 3700 San Martin
  Drive, Baltimore, MD 21218, USA} 
\altaffiltext{4}{Jansky Fellow; Department of Physics and Astronomy,
  Michigan State University, East Lansing, MI 48824, USA} 
\altaffiltext{5}{Clay Fellow}
\altaffiltext{6}{
Astronomy Department,
University of Illinois at Urbana-Champaign,
1002 West Green Street,
Urbana, IL 61801 USA
}

\altaffiltext{7}{
Department of Physics,
University of Illinois at Urbana-Champaign,
1110 West Green Street,
Urbana, IL 61801 USA
}
\altaffiltext{8}{Institute for Astronomy, University of Hawaii, 2680
  Woodlawn Drive, Honolulu HI 96822, USA} 
\altaffiltext{9}{Institute for Astronomy, University of Edinburgh
  Scottish Universities Physics Alliance, Royal Observatory, Blackford
  Hill, Edinburgh EH9 3HJ, UK}
\altaffiltext{10}{Department of Astronomy, University of Texas at
  Austin, Austin, TX 78712, USA} 
\altaffiltext{11}{Department of Physics and Astronomy, Johns Hopkins
  University, 3400 North Charles Street, Baltimore, MD 21218, USA} 
\altaffiltext{12}{Gemini Observatory, 670 North Aohoku Place, Hilo, HI
  96720, USA}
\altaffiltext{13}{Astrophysics Research Centre, School of Mathematics
  and Physics, Queen's University Belfast, Belfast, BT7 1NN, UK} 
\altaffiltext{14}{Department of Physics, Harvard University, 17 Oxford
  Street, Cambridge, MA 02138, USA}
\altaffiltext{15}{Astronomy Department, California Institute of
  Technology, MC 249-17, 1200 East California Boulevard, Pasadena, CA
  91125, USA} 
\altaffiltext{16}{Department of Astrophysical Sciences, Princeton
  University, Princeton, NJ 08544, USA}

\begin{abstract}
We present the \PS\ discovery of the long-lived and blue transient
\af, which was also detected by \galex\ with  coordinated observations
in the near-ultraviolet (NUV) band.  \af\ is associated with
the nucleus of an early-type galaxy at redshift $z$=0.4046 that
exhibits no evidence for star formation or AGN activity.  Four epochs
of spectroscopy reveal a pair of transient broad absorption features
in the UV on otherwise featureless spectra.  Despite the superficial
similarity of these features to P-Cygni absorptions of supernovae
(SNe), we conclude 
that \af\ is not consistent with the properties of known types of SNe.
Blackbody fits to the spectral energy distribution 
are inconsistent with the cooling, expanding ejecta of a SN,
and the velocities of the absorption features are too high to
represent material in homologous expansion near a SN
photosphere.  However, the constant blue colors and slow evolution of
the luminosity are similar to previous optically-selected tidal
disruption events (TDEs).  The shape of the optical light curve is
consistent with models for TDEs, but the minimum accreted mass
necessary to power the observed luminosity is only $\sim$0.002~\msun,
which points to a 
partial disruption model. A full disruption model predicts higher
bolometric luminosities, which would require most of the radiation to
be emitted in a separate component at high energies where we lack
observations. In addition, 
the observed temperature is lower than that predicted by pure
accretion disk models for TDEs and requires reprocessing to a
constant, lower temperature.  
Three deep non-detections in the radio
with the VLA over the first two years after the event set strict
limits on the production of any relativistic outflow comparable to
{\it Swift} J1644+57, even if off-axis. 
\end{abstract}

\section{Introduction}

When a star of mass \mstar\ and radius $R_{\star}$ has an orbit with a
pericenter passage sufficiently close to a black hole of mass \mbh, as
in the nucleus of a galaxy, it can be torn apart by tidal forces
\citep{hills}.  Approximately half of the debris of the star becomes
unbound and leaves the system on hyperbolic orbits, while the other
half remains bound to the black hole on parabolic orbits.  When the
bound material returns to pericenter, it can then accrete onto the
black hole and produce an optical transient \citep{rees}, called a
tidal disruption event (TDE).  TDEs are signatures of the presence of
otherwise quiescent black holes and, with better understanding of the
relevant physics, may prove useful in the long run both as
probes of black hole masses in distant galaxies and for the study of
accretion processes in a different regime than that of active galactic
nuclei (AGN).

The condition for disruption is that the pericenter distance of the
orbit is less than the tidal radius,
$r_t$=$R_{\star}$(\mbh/\mstar)$^{1/3}$.  For solar-type stars
disrupted by 10$^6$~\msun\ black holes, the characteristic blackbody
(BB) temperature (\tbb) for a black hole accreting at the Eddington
rate at this radius is $\sim$2.5$\times$10$^5$~K (e.g.,
\citealt{ulmer99,strubbe09}).  The radiation output is therefore
expected to peak in the extreme-ultraviolet (UV) and X-ray bands.
The characteristic light curve behavior prediction for TDEs involves a
rapid rise to maximum light, with a decline after the peak that falls
as $t^{-5/3}$ \citep{rees,ek89}. 

However, more detailed modeling has shown that actual TDEs should
exhibit more complex behavior.  The derivation of the $t^{-5/3}$ light
curve relies on the assumption that the spread of specific energy
with mass for the stellar debris is constant, but recent work has
shown that the internal structure of the star can modify these
expectations \citep{rrr,lodato09,stone,grr13}, with faster decline
rates 
predicted immediately after the peak.  Furthermore, most studies have
concentrated on complete stellar disruptions at the tidal radius,
while real disruptions can occur at a range of pericenter distances
and partial disruptions outside of the nominal tidal radius may also
contribute to the flare rate \citep{grr13}.   Finally, the conversion
of accreting mass, \mdot, to observable radiation is
not a simple process.  Models of the spectral energy distributions
(SED) of TDEs and their evolution have grown increasingly complex over
time, starting with thin disk models and adding thick disks and
outflows or winds to model super-Eddington accretion
\citep{lu97,ulmer99,strubbe09,strubbe11,lr11,james10jh}. 

Two relativistic TDEs, {\it Swift} J164449.3+573451 (\swone;
\citealt{bloom,levan11,burrows,baz11}) and {\it Swift} J2058.4+0516
(\swtwo; \citealt{swift2}) have been discovered on the basis of
$\gamma$-ray triggers. These objects have X-ray light curves
that approximately match the $t^{-5/3}$ decline rate, and appear
to have launched relativistic jets along the line of sight
\citep{gm11}, adding another potential emission component to the SED. 

Most of the TDE candidates reported in the literature are large
amplitude soft X-ray flares from galaxy nuclei (e.g.,
\citealt{bade,komossa99,li02,komossa04,esquej,maksym13}).  
These generally have poorly sampled light curves, but have the
predicted soft spectra and light curve decay rates that are
consistent with a $t^{-5/3}$ decline, for suitable assumptions about
the time of disruption.
\citet{gezari09} presented \galex\ observations of three TDE
candidates with SEDs that had \tbb\ $\gtrsim$5$\times$10$^4$~K and
light curves exhibiting evidence of  $t^{-5/3}$ declines.  \citet{iya}
interpreted the fast fading and luminous nuclear transient PTF~10iya
as the early super-Eddington phase of accretion in a TDE.

\citet{vv11} discovered the first two optically-selected TDE
candidates in repeated Sloan Digital Sky Survey (SDSS) imaging of
Stripe 82, which we denote \tone\ and \ttwo.  These transients were
selected on the basis of their unusually blue colors and slow
evolution, which made them stand out from normal supernovae (SNe) and
AGN variability.  Subsequently, \citet{10jh} described the Pan-STARRS1
(PS1) discovery
of \jh, the first TDE with a well-sampled optical light curve on both
the rise and decline from maximum light, which makes it a benchmark
object for studies of the TDE process (e.g.,
\citealt{james10jh,bog13}).  The spectra 
of \ttwo\ exhibited broad H$\alpha$ emission, while \jh\ had broad
\ion{He}{2} emission lines.  These were the first definitive detections
of spectral features from TDEs, although some galaxies in the
literature with unusual nuclear spectra have been claimed to be
produced by the late-time effects of TDEs
\citep{sb95,bog04,komossa08,wang12,yang13}. 

In this work, we describe the discovery of a new TDE by the PS1
survey, \af, which is the first TDE to exhibit broad UV absorption
features.  We give a description of the optical, 
UV, and radio observations in Section 2.  In Section 3, we analyze the
host galaxy and set upper limits on star formation and AGN activity.
We describe the evolution of the SED
and isolate the spectrum of the transient in Section 4.  In Section 5,
we examine a SN interpretation for the properties of
\af\ and find that it cannot be fit by any known model.  We then 
interpret \af\ as a TDE in Section 6 before concluding in Section 7.
Throughout this paper, we adopt the flat $\Lambda$CDM {\it
  Planck}+{\it WMAP}+high-$\ell$+BAO
cosmology of \citet{planck} with  $H_0$=68~\kms~Mpc$^{-1}$,
$\Omega_{\mathrm{M}}$=0.31, and $\Omega_{\Lambda}$=0.69.
All quoted dates are UT, and all magnitudes are reported on the AB
scale.

\section{Observations}

\subsection{Discovery and Photometry}
\label{sec:phot}

The PS1 telescope has a 1.8~m diameter primary mirror that images a
field with a diameter of 3.3\degr\
\citep{PS1_optics} onto a total of 60 $4800\times4800$ pixel
detectors, with a pixel scale of 0.258\arcsec\ \citep{PS1_GPCA}.
A more complete description of the PS1 system, hardware
and software, is provided by \cite{PS1_system}. 
The nightly PS1 Medium Deep Survey (MDS) observations are obtained
through a set of five broadband filters, designated as \gps, \rps,
\ips, \zps, and \yps, with a typical cadence of 3~d between
observations in \griz.
  Although the filter system for PS1 has much in
common with that used in previous surveys, such as the SDSS
\citep{SDSS}, there are differences, with
further information on the passband shapes described
by \cite{PS_lasercal}. Photometry is reported in the ``natural'' PS1 
system, $m=-2.5\log(f_{\nu})+m'$, with a single zeropoint adjustment $m'$
made in each band to conform to the AB magnitude scale
\citep{JTphoto}.  PS1 magnitudes are interpreted as
being at the top of the atmosphere, with 1.2 airmasses of atmospheric
attenuation being included in the system response function.

 PS1 data are processed 
through the Image Processing Pipeline (IPP; \citealt{PS1_IPP}) on a
computer cluster at the Maui High Performance Computer Center. The
pipeline runs the images through a succession of stages,  
including flat-fielding (``de-trending''), a flux-conserving warping
to a sky-based image plane, masking and artifact removal, and object
detection and photometry.  Transient detection using IPP photometry is
carried out at Queen’s University Belfast. Independently, difference
images are produced from the stacked nightly MDS images by the
\texttt{photpipe} pipeline \citep{rest05} running on the Odyssey
computer cluster at Harvard University. The discovery and data
presented here are from the \texttt{photpipe} analysis.

\begin{figure*}
\epsscale{1.2}
\plotone{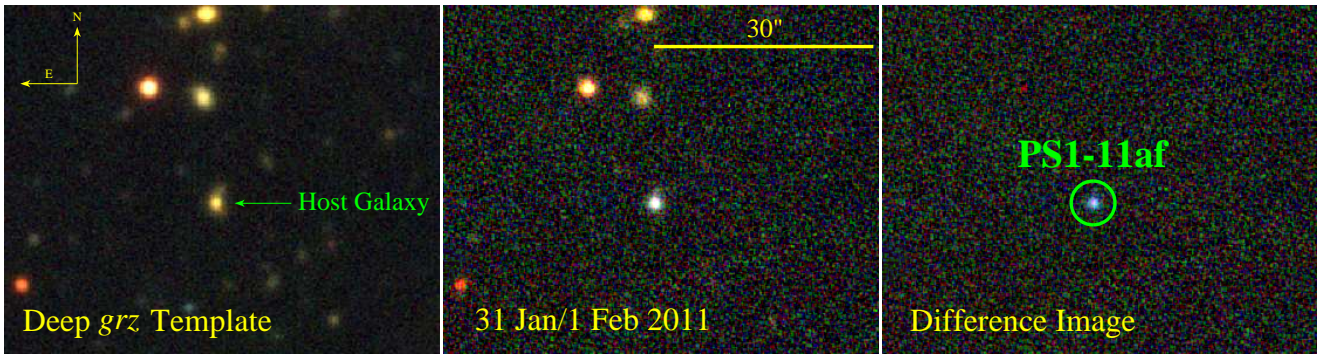}
\caption{False-color images of the field of \af, with
  \gps\rps\zps\ mapped to blue, green, and red, respectively.  The
  logarithmic color scales are the same in each panel.  The left panel
  displays the deep template images created from pre-outburst PS1
  data.  The image in the middle panel was formed from images taken on
  consecutive nights shortly after peak of the outburst.  The right
  panel is the difference image formed by subtraction of the template
  images on the left from the images in the middle.  Note the strong
  blue color of the transient relative to its host galaxy and other
  field sources. 
}
\label{rgbfig}
\end{figure*}

We first detected \af\ in \gps\rps\ images obtained on the night of
2010 December 30  after non-detections in
our first observations of that field for the observing season on 2010
December 15/16 (\gps\rps\ips).  The transient rose to a peak in late
January of 2011 and slowly faded thereafter, remaining detectable in
PS1 imaging when observations ceased at the end of April due to solar
conjunction.  The mean position of \af\ was 
$\alpha$ = 9$^{\mathrm  h}$57$^{\mathrm m}$26\fs815, 
$\delta$ = $+$03\degr14$'$01\farcs00 (J2000), with an uncertainty of
0$\farcs$1 in each coordinate. 
  False-color images of \af\ near maximum light and its
host galaxy are shown in Figure~\ref{rgbfig}.  The blue color of the
transient relative to its red host galaxy is immediately apparent.

\begin{figure}
\epsscale{1.2}
\plotone{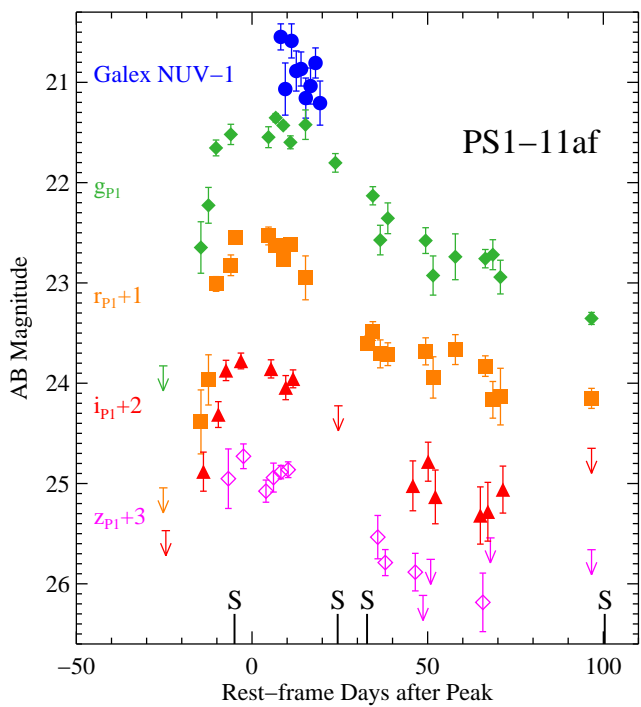}
\caption{Light curve of \af.  The blue circles are from the
  \galex\ NUV band, with the \griz\ photometry below, and the
  constant offsets are noted.  Vertical lines marked with an ``S''
  show the dates of the spectroscopic observations.
}
\label{lcfig}
\end{figure}

We construct deep template observations from the pre-outburst images
and subtract them from the PS1 observations using \texttt{photpipe}
\citep{rest05}.  Details of the photometry and generation of PS1
transient light curves are given by \citet{rest} and \citet{scolnic}.
The final \af\ photometry, after correction 
for $E(B-V)=0.027$~mag of Galactic extinction \citep{eddiedoug,sfd98}, is
given in Table~\ref{phottab} and shown in Figure~\ref{lcfig}.
Fourth-order polynomial fits to the \gps\ and \rps\ light curves near
peak give dates for maximum light at Modified Julian Dates (MJDs) of
55581$\pm$2 days (=2011 January 20), which we adopt as the time of
peak throughout this paper.

The field of \af\ was monitored in the near-ultraviolet band (NUV;
\citealt{morrissey07}) as part of the \galex\ Time Domain Survey (TDS; 
\citealt{suvi13}) from 2011 January 31 to 2011 February 16.  The
transient was detected in each observation during this period,
while nothing was detected at the position of \af\ in either of the
NUV or far-ultraviolet (FUV) bands in prior \galex\ observations
in 2008 and 2010.  We perform aperture photometry of \af\ during
outburst and list the results in Table~\ref{phottab}.  Further details
of the \galex\ survey design and data products were presented by
\citet{suvi13}.

In addition, we observed the host of \af\ on 2013 May 20 in $J$
and $H$ for a total of 1200 and 1254~s, respectively, using the
FourStar Infrared Camera \citep{fourstar} on the 6.5~m Magellan Baade 
telescope.  The images were flat fielded, sky subtracted, and stacked
with standard tasks in IRAF\footnote{IRAF is 
  distributed by the National Optical Astronomy Observatories, 
    which are operated by the Association of Universities for Research
    in Astronomy, Inc., under cooperative agreement with the National
    Science Foundation.} using the instrument pipeline.  The
photometry was calibrated using 2MASS stars in the field.

\subsection{Spectroscopy}
\label{sec:spec}

We obtained four epochs of spectroscopy of \af\ in outburst between
2011 January 13 and 2011 June 10 using the Low Dispersion Survey
Spectrograph-3 (LDSS3; an updated verion of LDSS-2, \citealt{ldss})
on the 6.5~m Magellan Clay telescope, the Blue 
Channel Spectrograph (BC; \citealt{schmidt89}) on the 6.5~m MMT, the
Inamori-Magellan Areal Camera and Spectrograph 
(IMACS; \citealt{imacspaper}) on the Magellan Baade telescope,
and the Gemini Multi-Object Spectrograph (GMOS; \citealt{hookgmos}) on
the 8~m Gemini-South telescope. A full log with
details of the observations is given in Table~\ref{spectab}.
Conditions were clear for the initial LDSS3 and BC observations,
but poor and variable seeing compromised the IMACS spectra.  We
combine the IMACS spectra on consecutive nights to increase the
signal-to-noise ratio.

After \af\ had faded away, we returned to obtain spectroscopy of the
host galaxy using similar 
BC and LDSS3 setups as those used for the original observations,
including the same slit position angles (PAs) to within 5\degr.  We
also obtained spectroscopy with LDSS3 using a significantly redder
setup, designed to cover H$\alpha$ at the rest frame of the host
galaxy.  A few additional host spectra were obtained at
PA=$-10$\degr\ for reasons discussed below, but only
when that angle was close to the parallactic angle.

We also obtained several images as part of the acquisition
process using LDSS3, IMACS, and GMOS.  We process the two-dimensional
frames using standard tasks in IRAF.  Multiple attempts to obtain the
GMOS spectrum over the course of several days were aborted due to bad
weather so we only obtained a $g'r'i'z'$ acquisition sequence on each
date.  We stack the images in each filter and report the average
dates in Table~\ref{phottab}.  We subtract the PS1 templates in the
appropriate filters from the acquisition images and calibrate the
photometry to PS1 stars in the field.

\begin{figure}
\epsscale{1.2}
\plotone{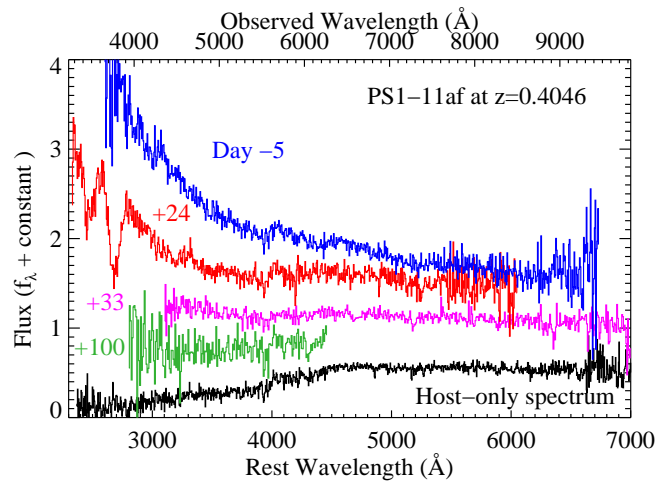}
\caption{Observed spectra of \af\ with constant flux offsets for
  clarity.  The spectra are labeled with the number of rest-frame days
  relative to peak.  Note the blue colors at early times and the two
  deep UV absorption features on Day $+$24.
  The bottom spectrum (black) shows the combined spectrum of the host
  galaxy after the transient faded.
}
\label{allspecfig}
\end{figure}

We reduce the spectroscopic data using standard tasks in IRAF
along with our own IDL procedures to apply a
flux calibration and correct for telluric absorption.  The longslits
were always aligned within 10\degr\ of the parallactic angle
to mitigate the possible effects of differential atmospheric
dispersion \citep{fil82}.  We did not use order-blocking filters, so
second-order light contamination is a concern with a source this
blue.  Our early LDSS3 observation was taken using the standard
0\farcs75 longslit located 2$'$ blueward of the center of the field
combined with the VPH-all grism.  Past experience, along with tests
using order-blocking filters on blue standard stars, has demonstrated
that this instrument setup combination suffers very little
contamination from second-order light.  More generally, we use
observations of both relatively blue (sdO spectral types) and red
(F-type) standard stars taken both with and without order-blocking
filters to define the flux calibrations for the BC and Magellan
spectra.  Despite our best efforts, we caution that some contamination
may still be present at observed wavelengths longward of
$\sim$8000~\AA.  For GMOS-S, we use archival observations of EG21 to
define the flux calibration.  The final spectra are presented in
Figure~\ref{allspecfig}.  All spectra exhibit absorption from the
\ion{Ca}{2} H+K doublet at redshift $z=0.405$, which we refine in
Section~\ref{sec:sfr} to $z=0.4046$ and adopt as the host redshift
throughout this paper. 

\subsection{Radio Observations}

We observed the field of PS1-11af with the Karl G. Jansky Very Large Array
\citep{dp10} three times.  
The first epoch was 2011 March 29.03 (Project 10A-214; PI:
Soderberg), and subsequent epochs were obtained beginning 2012 January
7.38 (Project 11B-192; PI: Chomiuk) and 2013 May 31.96 (Project
13A-437; PI: Soderberg). 
Our observations are summarized in Table~\ref{radtab}.
The first observation was conducted with the old VLA
system with two 128 MHz windows centered at 4.8 and 5.0 GHz.
The latter two observations were conducted with the new WIDAR
correlator \citep{evla} at
a mean frequency of 5.9 GHz (lower sideband frequency
centered at 5.0 GHz; upper sideband frequency centered at 6.75 GHz). 

In all epochs, we use standard data reduction procedures in AIPS
\citep{aips}.  We use 3C286 for bandpass and flux calibration, and
J1024$-$0052 for gain calibration.  We flag and excise channels
affected by radio frequency interference, which results in an
effective bandwidth of 
$\sim$1.7~GHz.  A $\sim$35 mJy source 4$^\prime$ from the position of
\af\ is used for self-calibration.
We do not detect significant radio emission from \af\ in any epoch and
set 3$\sigma$ limits of 51, 30, and 45~$\mu$Jy, respectively, from the
root-mean-squared (RMS) flux values of the images.

\section{Host Galaxy}

The host galaxy of \af\ is shown in the left panel of 
Figure~\ref{rgbfig}.  It is apparent that the host is relatively
red and the core is rather symmetric, indicative of a bulge-dominated
or early-type galaxy.  The host is marginally resolved in the PS1
template images (seeing full-width at half-maximum,
FWHM$\approx$1.1\arcsec).  An asymmetric extension of
flux to the north-northwest (PA$\approx$$-10$\degr) is visible.  In
our images obtained in the 
best seeing (with LDSS3 in $g'$ and FourStar), this extension appears
somewhat offset from the main part of the galaxy.  The offset
($\sim$2$\arcsec$) is 11~kpc at this redshift.  It is not immediately
apparent whether this represents substructure within the host galaxy
or possibly a companion galaxy.

We use the PS1 image with our highest-significance detection of
\af\ in \rps\ (MJD=55590.5) to perform relative astrometry between the
deep template image and the detection of the source.  Forty common
point sources were used to tie the two images together, with a
RMS dispersion of 28 milliarcsec in each coordinate.  The
offset between the detection of \af\ and the core of
its host galaxy in the template is 60$\pm$62 milliarcsec, including
the uncertainties in the object and nuclear centroids, consistent
with \af\ being a nuclear event.  However, the 3$\sigma$ uncertainty
corresponds to 1~kpc, which contains a significant fraction of the
stars in this compact host galaxy.

We collect photometry of the host galaxy of \af\ using the deep PS1
template images and FourStar data described in
Section~\ref{sec:phot}.  Because of the ambiguous origin of the
flux extension, we report 
aperture photometry in Table~\ref{hosttab} using apertures of radius
1\farcs15 and 3\farcs0.  The narrower aperture captures the main core
of the galaxy (possibly the bulge), while the wider one includes the
extension plus the 
wings of the central galaxy.  The FourStar data were taken under
significantly better seeing (FWHM$\approx$0\farcs75) than the PS1
images.  The core of the host is not well resolved, so the quoted
aperture photometry was calculated after convolving the FourStar data
with a Gaussian to approximately match the seeing in the PS1 templates
and make the narrow apertures more directly comparable between
instruments. We stack the pre-outburst
\galex\ photometry and set upper limits on the host galaxy flux in the
UV. 

\begin{figure*}
\epsscale{1.2}
\plotone{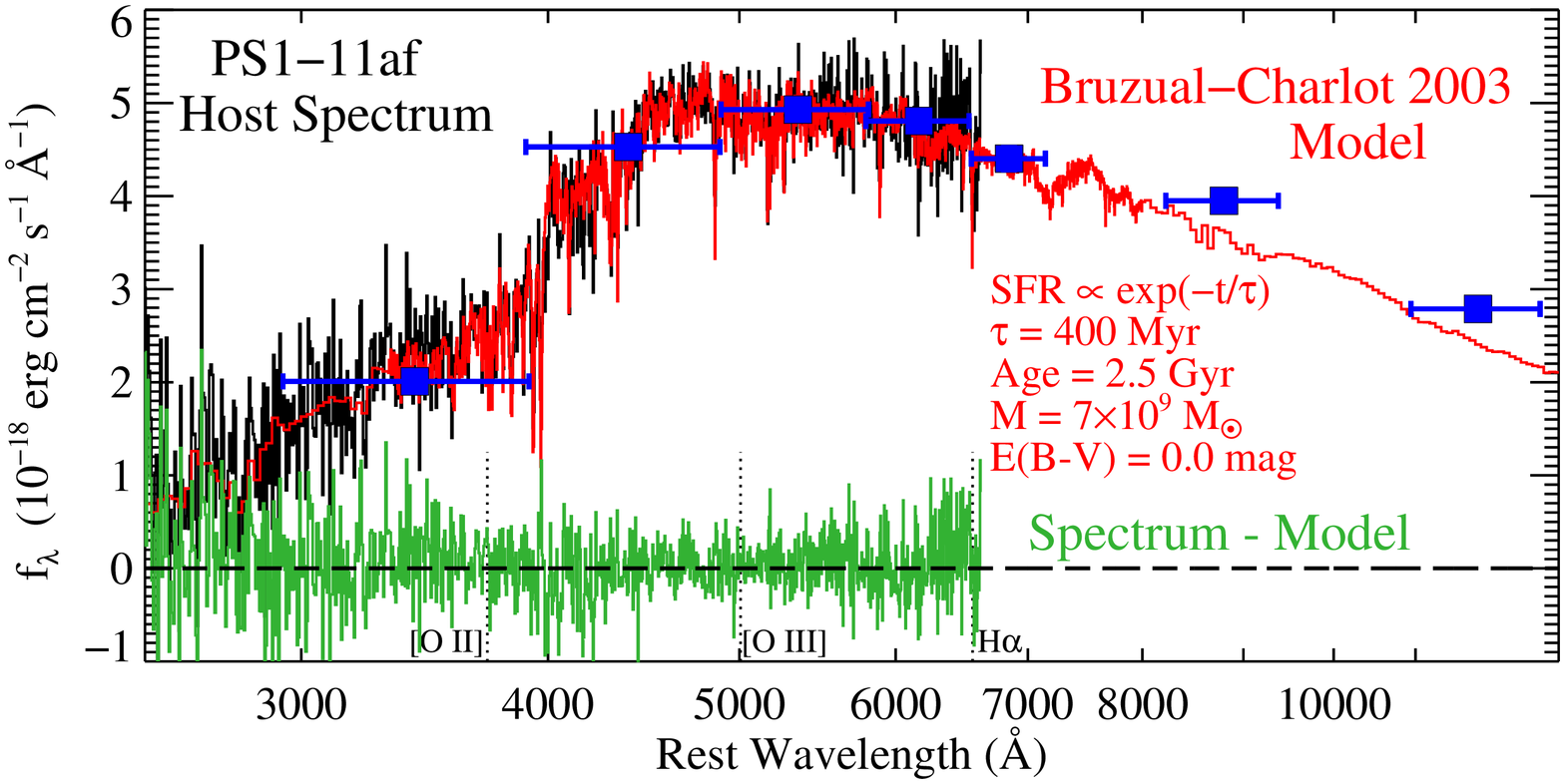}
\caption{Host galaxy spectrum of \af\ (black) compared to a best-fit
  spectral synthesis model (red).  The parameters for the \citet{bc03}
  spectral model are given in red.  The \grizy\ PS1 and $JH$ FourStar
  photometry points are plotted as the blue squares.  The green
  spectrum is the difference between the observed spectrum and the model.
  The model has an age of 2.5~Gyr and is an excellent fit to the
  data.  The dotted vertical lines mark the expected wavelengths of
  [\ion{O}{2}] $\lambda$3727, [\ion{O}{3}] $\lambda$5007, and
  H$\alpha$, which would be signatures of either star formation or AGN
  activity.  No emission from any line is detected.
}
\label{hostfig}
\end{figure*}

In addition to the photometry, we have obtained about 4 hours of
spectroscopy of the host galaxy, as described in
Section~\ref{sec:spec}.  The resolutions of the BC and LDSS3 spectra
are similar, so we combined all of the spectra by rebinning to a
common wavelength scale and performing a weighted average over the
overlap regions.  We plot this combined spectrum in
Figure~\ref{hostfig}, after scaling to the host photometry in the
narrow apertures because they more closely match the size of the
longslit spectroscopic apertures.  

\subsection{Limits on Current Star Formation}
\label{sec:sfr}

It is obvious from the strength of the 4000~\AA\ break that the
spectrum shown in Figure~\ref{hostfig} is that of an older stellar
population.  The Balmer lines are only present in absorption and there
are no emission lines of any type, while absorption lines from several
metals are strong.  We cross-correlate this spectrum with the single
stellar population age models of \citet{bc03}.  All models that are
good fits have ages $>$1~Gyr, with the best fit at solar metallicity
being 2.5~Gyr. The best-fit cross correlations give a precise
redshift of $z=0.4046 \pm 0.0001$.

We use the FAST program of \citet{mariska} to simultaneously fit the
spectrum and photometry to a suite of solar metallicity stellar
population models from \citet{bc03} with more complicated star
formation histories.  The best fit is overplotted in
red, and uses an exponentially-declining star formation rate (SFR)
with an e-folding timescale of 400~Myr and an age of 2.5~Gyr.    Only
the \rps\ and \ips\ points were used to scale the spectrum, so the
good agreement in the other bands is reassuring.  The total stellar
mass is 7$\times$10$^{9}$~\msun\ when scaled just to the photometry in
the inner aperture. The photometry in the outer aperture is
$\sim$0.6~mag brighter, so if all of the flux is assigned to the host,
then the total stellar mass is closer to 1.2$\times$10$^{10}$~\msun.

The difference spectrum shown at the bottom of Figure~\ref{hostfig}
shows no strong deviations between the model and the observed
spectrum.  In particular, no emission lines are visible.  We use the
difference spectrum and observed errors to set
3$\sigma$ upper limits on the ([\ion{O}{2}] $\lambda$3727, H$\alpha$)
emission fluxes of (5.2, 4.8)$\times 10^{-18}$~erg~cm$^{-2}$~s$^{-1}$ in
10~\AA\ bins around the rest wavelengths of the lines.  According to
the calibration of \citet{kennicutt98}, these correspond to upper
limits on the SFR of the host galaxy of (0.04, 0.02)~\msun~yr$^{-1}$.
Similarly, we also use the non-detections in the UV by \galex\ and the
calibrations of \citet{kennicutt98} to set 5$\sigma$ upper limits on
the SFR of 0.8~\msun~yr$^{-1}$.  Combined, these indicate that there
is no sign of star formation or any young stellar population in the
host galaxy at the position of \af.  In
addition, we inspect our spectra of the host galaxy obtained at the
PA of $-10$\degr.  While the surface brightness of the extension is
too low to obtain spectra with a good S/N in our limited exposure
time, inspection of the two-dimensional frames shows no obvious
emission lines.

\subsection{Evidence Against an AGN Host}

An outburst in a persistently accreting AGN could masquerade as a true
transient.  However, the non-detection in the UV by \galex\ argues
strongly against the presence of a luminous unobscured AGN in the
nucleus.  The stacked non-detections in the \galex\ bands correspond
to absolute magnitudes $M_{\mathrm{UV}}$$>$$-17$~mag or UV continuum
luminosities $\nu L_{\nu}$$\lesssim$$6\times$10$^{42}$~erg~s$^{-1}$.
Also, the difference spectrum at the bottom of Figure~\ref{hostfig}
exhibits no emission lines from the broad-line region (BLR) of an AGN.
The spectra of both obscured and unobscured AGN could also manifest
strong forbidden 
emission lines from a narrow-line region (NLR), but those are not
present either. The 3$\sigma$ limit on the [\ion{O}{3}] $\lambda$5007
luminosity is $<$1.4$\times$10$^{39}$~erg~s$^{-1}$, which falls below
any of the optically-selected type II AGNs at similar redshifts in the
zCOSMOS survey \citep{bongiorno}.  However, we cannot exclude the
possibility of a low-luminosity AGN, as samples in the local universe
extend to significantly lower emission-line luminosities than our
limits for the host of \af\ \citep{ho08,hao05}. 
All of these upper limits would be made weaker by the presence of strong
dust obscuration in the nucleus, but the blue color of the transient
(Figure~\ref{allspecfig}; see below) is incompatible with a large dust
column along the line of sight.

The deep non-detection of the host galaxy in the UV prior to the
detection of \af\ also
represents a strong argument against normal accretion rate
fluctuations in an AGN.  As described by \citet{10jh}, quasars and AGN
exhibit variability of a lower amplitude in \galex\ observations than
the $>$2.5~mag increase in brightness of \af\ relative to quiescence.
In the full \galex\ TDS observations \citep{suvi13}, most sources
with such large amplitude NUV flares are classified as cataclysmic
variables or M-dwarf flares.  The only extragalactic sources having
similarly large amplitude flares after non-detections in quiescence
were the TDE PS1-10jh and a few SNe \citep{10jh}.

Another possibility is that \af\ is an outburst from a blazar
\citep{urry}, but we 
would expect to detect a blazar in the radio.  Near the peak of the
outburst,  \af\ had a $V$-band magnitude of $\sim$21.5, corresponding
to $\nu L_{\nu}$$\approx$3$\times$10$^{43}$~erg~s$^{-1}$.  The unified
blazar SEDs of \citet{fossati} predict radio fluxes in the range 
0.4--15 mJy for different ranges of radio loudness.  Even the faintest
of these is $\sim$8 times brighter than our 3$\sigma$ upper
limit in the first epoch.  The broad UV absorption features of
\af\ would also be unexpected in the synchrotron-dominated spectra of
a blazar.

\begin{figure}
\epsscale{1.2}
\plotone{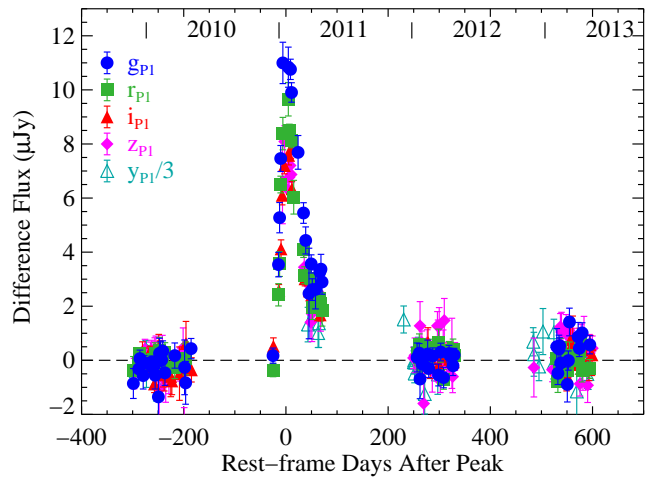}
\caption{Long-term difference flux light curve of \af\ in five
  filters.  The calendar years of observations are labeled across the
  top.  The \yps\ light curve has
  been divided by 3 for clarity because the observations are shallower
  than in the other bands so the typical scatter is larger.
  This light curve was constructed using a template image composed of
  an average of several of these same epochs of observation, which
  means that the zeropoint of the flux scale has an arbitrary constant
  offset level that we set to zero in the first year.
}
\label{alltfig}
\end{figure}

As a final check against AGN-like variability, we present difference
light curves in five filters for \af\ from all four years of PS1
observations obtained to date in
Figure~\ref{alltfig}.
This light curve shows no sign of variability or any long-term
trend during the other three observing seasons to date.
The host galaxy was also observed by
SDSS a decade prior to our detection of \af\ (object ID: SDSS
J095726.82$+$031400.9) and the model magnitudes reported by
\citet{SDSS} from 2001 February 20 are consistent within 1$\sigma$
of our PS1 photometry in the large aperture (Table~\ref{hosttab}). 

The fast rise in \gps\ of \af\ is especially notable in this context.
The host flux in the narrow aperture was \gps=23.42~mag, but \af\ rose
on a timescale of less than a month to a peak of
\gps$\approx$21.5~mag, implying a rise of more than 2~mag (the
quiescent flux is clearly dominated by star light;
Figure~\ref{hostfig}).  The characteristic amplitude of quasar
photometric variability on these timescales is only $\sim$0.1~mag,
with the amplitude of variability increasing on longer timescales
(e.g., \citealt{webb,cmacleod12}).  This is very different from the
long-term \af\ light curve, which is consistent with a constant in the
observing seasons lacking the transient.  

\section{Evolution of \af\ }
\label{sec:evo}

The light curve of \af\ exhibits a relatively
rapid rise of $\sim$2~mag from the first detections in \gps\rps\ to
the peak 14.5 rest-frame days later.  The subsequent decline from
maximum light was much slower, with a decline of $\lesssim$2~mag by
the time of our final GMOS observations at nearly Day $+$100 (this
notation refers to the phase of the object in rest-frame days relative
to maximum light).  This basic light curve shape was present in all
filters (Figure~\ref{lcfig}), which implies very little color
evolution. 

\begin{figure}
\epsscale{1.2}
\plotone{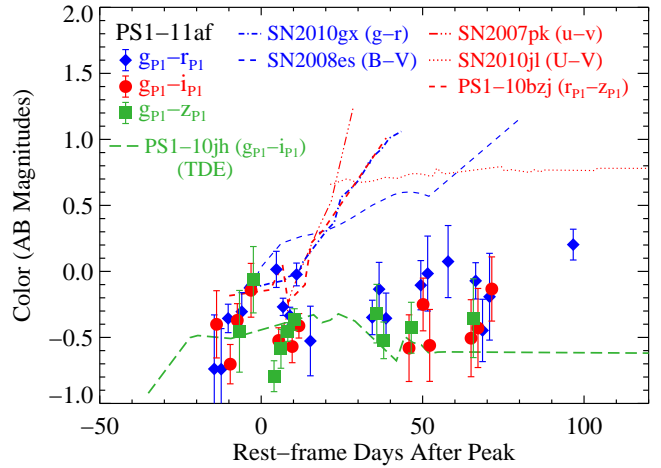}
\caption{Observed color evolution of \af\ compared to several SNe and
  one TDE.  The comparison objects are labeled with the observed
  filters used for the color curves.  Note the blue and relatively
  constant colors of \af\ compared to the SNe, which rapidly evolve
  redward. 
  The objects and the rest-frame
  wavelengths of the plotted filters are: SN~2010gx/PTF~10cwr ($g$,
  3770~\AA; $r$, 4975~\AA; \citealt{10gx,quimby11}), SN~2008es ($B$,
  3600~\AA; $V$, 4490~\AA; \citealt{miller08es,gezari08es}), SN~2007pk
  ($u$, 3465~\AA; $v$, 5468~\AA; \citealt{07pk}), SN~2010jl ($U$,
  3430~\AA; $V$,  5410~\AA; \citealt{zhang10jl}), PS1-10bzj (\rps,
  3740~\AA; \zps, 5250~\AA; \citealt{10bzj}), and PS1-10jh (\gps,
  4110~\AA; \ips, 6430~\AA; \citealt{10jh}). 
}
\label{colcompfig}
\end{figure}

The weak evolution of the rather blue SED is one of the most important
clues to the nature of \af, so we 
examine it in detail.  We present the observed colors in the PS1
filters in Figure~\ref{colcompfig}.  We calculate \gr\ data
points only when we have \gps\ and \rps\ observations on the same
night.  The other filters were generally not obtained simultaneously,
so we interpolate the less-noisy \gps\ light curve to the dates of
observation of the \ips\ and \zps\ data using low-order polynomials.
We determine the uncertainties in the derived colors by repeatedly
re-fitting the observed data points after adjusting by Monte Carlo
resamplings of the errors.  We also apply this procedure below
whenever we need to interpolate the photometry to common epochs.

The \gi\ and \gz\ data points in Figure~\ref{colcompfig} are
consistent with constant values from Day $-$10 to $+$70.  The
\gr\ color is similarly constant over this time interval, but the
final point near Day $+$97 is redder and there is a hint of
a bluer color at earlier times.  We compare the
\af\ data to several SNe as well as the TDE PS1-10jh \citep{10jh}.
The color curves for the comparison objects are color coded so that
lines of a given color correspond to approximately similar rest-frame
wavelengths as the \af\ data points of the same color.  This largely
eliminates the need to apply uncertain $K$-corrections.  For example,
\rps\ and \zps\ at the redshift of PS1-10bzj ($z=0.650$;
\citealt{10bzj}) are at rest-frame wavelengths of 3740 and 5250~\AA,
respectively, which are close to \gps\ and \ips\ at the redshift of
\af\ (cf. Table~\ref{hosttab}), so both are plotted in red.  The key
point is that regardless of any extinction or $K$-correction, the SNe
are only briefly as blue as \af\ and rapidly 
become redder with time, while the TDE PS1-10jh \citep{10jh} has a
similarly blue color that does not strongly evolve with time.  We
discuss the implications of this observation in more detail in
Section~\ref{sec:sn}. 

\begin{figure}
\epsscale{1.2}
\plotone{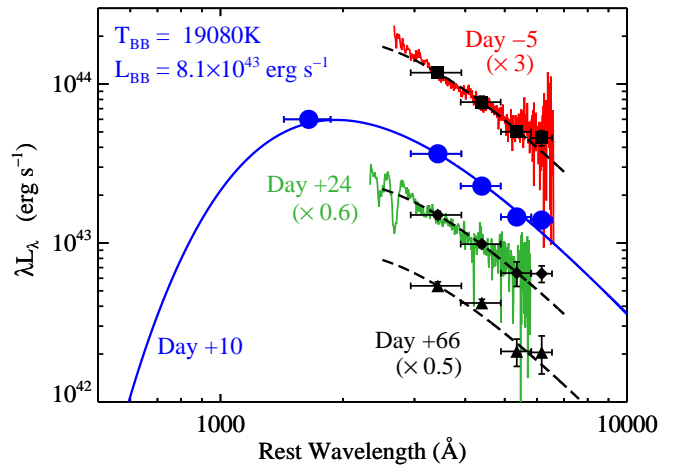}
\caption{SED evolution of \af.  The photometry has been
  interpolated to four common epochs.  The data points with blue
  circles come from Day $+$10, near the beginning of the
  \galex\ observations.  The other epochs have been offset for clarity
  by the multiplicative factors listed in parentheses. The dashed
  lines at the other three epochs represent the same BB fit as the
  thick blue line, but scaled in flux to
  match the \gps\ point at the appropriate epoch.  This BB fit
  continues to be a good fit for the data at other epochs, indicating
  a lack of strong SED evolution.  The red and green lines are the
  host galaxy subtracted spectra from our first two epochs of
  spectroscopy. 
}
\label{sedfig}
\end{figure}

We now investigate the SED of \af\ by interpolating the observed
photometry to four common epochs, which we display in
Figure~\ref{sedfig}. The \galex\ data provide an important constraint
on the SED, but are limited in phase coverage.  Therefore, we first
examine the SED on Day $+$10, near the beginning of the
\galex\ observations.  We fit a single-temperature BB to
the NUV through \zps\ SED and overplot the fit as the blue solid
line (Figure~\ref{sedfig}).  The best fit has \tbb=19,080$\pm$750~K,
with a luminosity (\lbb) of 8.1$\times$10$^{43}$~erg~s$^{-1}$.  In
addition, 
we plot the \griz\ photometry at the epochs of our first two spectra
(Days $-$5 and $+$24) and the date of our final \zps\ detection (Day
$+$66).  The solid dashed lines overplotted on the photometry for each
epoch are the same BB fit scaled to the \gps\ flux on each date.  The
same BB fit from the \galex\ epoch is also an adequate fit to the
\griz\ SEDs separated by 71 rest-frame days.

\begin{figure}
\epsscale{1.2}
\plotone{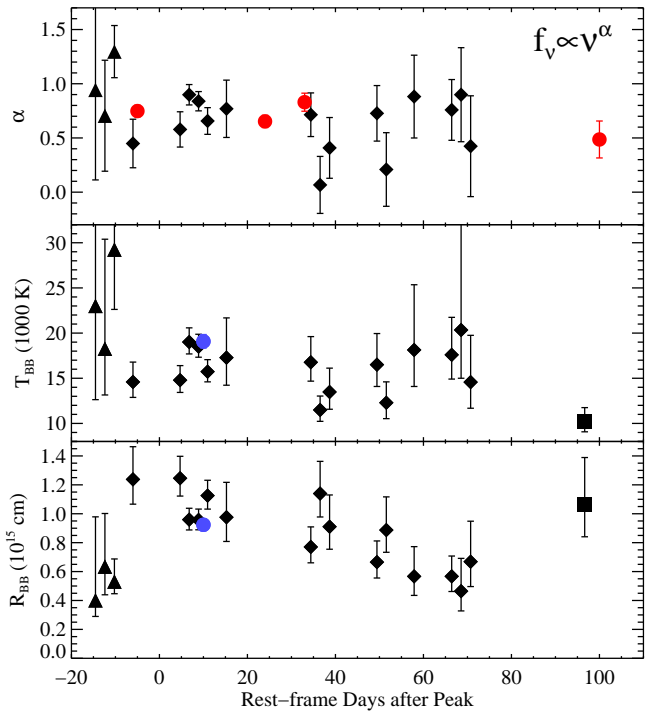}
\caption{Power law (top) and BB (middle, bottom) fits to the optical
  photometry  of \af.  The black
  diamonds mark fits to \griz, while triangles are fits to
  \gps\rps\ips\ (before the start of the \zps\ observations), and the
  square is a fit to the late-time $g'r'$ data only.  The blue 
  circles in the middle and bottom panels are from the BB fit to
  \galex$+$PS1 data shown 
  in Figure~\ref{sedfig}.  The red circles in the top panel are fits
  to the galaxy-subtracted spectra from Section~\ref{sec:galsub}.
}
\label{tempfig}
\end{figure}

Although the PS1 data are largely on the Rayleigh-Jeans tail of
the SED and we lack NUV observations for most of the light curve, we
still obtain useful constraints on the SED using our optical data.  
We interpolate the \ips\ and \zps\ light curves to the
dates on which we have both \gps\ and \rps\ data.  Our BB fits to the
available photometry are shown in the middle and bottom panels of
Figure~\ref{tempfig}.  As we inferred from the color evolution, the
best-fit \tbb\ is consistent with a constant from Days $-$10 to
$+$70.  The NUV+PS1 fit from Figure~\ref{sedfig} is shown as a
blue circle and is consistent with our optical-only fits.  

We have no {\it a priori} reason to believe that the SED of
\af\ should be well approximated by a single-temperature BB, so we
also fit the \griz\ photometry with power laws of the form
$f_{\nu}$$\propto$$\nu^{\alpha}$ and plot the best-fit power-law
indices in the top panel of Figure~\ref{tempfig}.  These fits again
demonstrate a lack of significant SED evolution during the course of
our observations.  A weighted average of the fits to all four
filters (diamonds in the top panel of Figure~\ref{tempfig}) gives
$\alpha=0.73\pm0.05$. 

This is notable for being significantly bluer
than AGN SEDs over this wavelength range.  Composite quasar templates
have mean values for $\alpha$ in the range $-0.32$ to $-0.50$ for the
NUV-optical continuum depending on the selection criteria
\citep{brotherton,vdb01}.  These values represent average quasar
SEDs, but \cite{wilhite05} isolated the variable component of the
spectra from multiepoch SDSS spectroscopy.  This variable component is
bluer than the average, but still has an NUV-optical slope of only
$\alpha$$\approx$0.  This is also not as blue as \af\ and provides
additional evidence that it is not a result of normal AGN
variability.  We note that \af\ is also bluer than the canonical
$\alpha$=$1/3$ value for multicolor BB disk emission \citep{pr72}.  In
a pure AGN accretion disk model, this implies that the SED of
\af\ does not include as much emission from the cooler large radii as
in normal AGN disks.  A simple explanation is that TDE accretion
disks are smaller than those of AGN, although below we discuss the
reasons to believe that the optical emission of \af\ does not directly
originate in a disk.

\subsection{Galaxy-subtracted Spectra}
\label{sec:galsub}

The observed spectra of \af\ shown in Figure~\ref{allspecfig} exhibit
\ion{Ca}{2} H$+$K absorption from the host galaxy as well as several
undulations that correspond to similar features in the host spectrum
(Figure~\ref{hostfig}).  A comparison of the host galaxy photometry
(Table~\ref{hosttab}) and the transient light curve
(Table~\ref{phottab}) reveals that even near maximum light we should
expect a significant contribution from the host galaxy to the observed
fluxes, particularly at the redder wavelengths.  We
now use the information about the host 
galaxy from Section~\ref{sec:sfr} and the photometric properties of
\af\ from Section~\ref{sec:evo} to subtract the host galaxy
contribution from the observed spectra and isolate the spectrum of the
transient. 

\begin{figure}
\epsscale{1.2}
\plotone{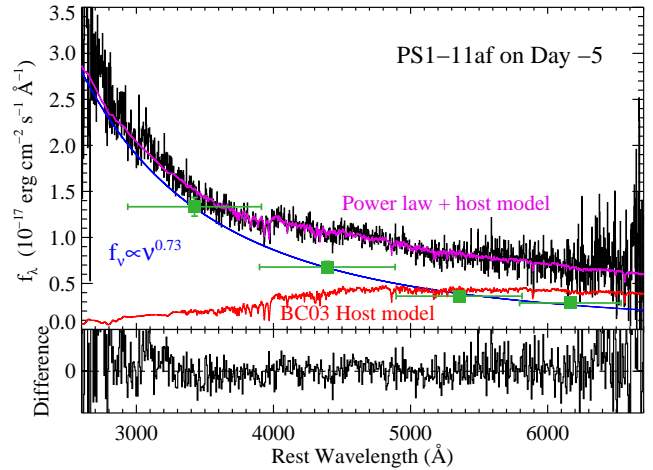}
\caption{Demonstration of our procedure for determining the proper
  amount of galaxy light to subtract from each epoch of spectroscopy.
  The observed spectrum (black) is modeled as the sum (magenta) of our
  best-fit host galaxy model (red; Figure~\ref{hostfig}) and a
  power-law continuum with a fixed slope (blue).  The difference
  between the  
  observed spectrum and the sum is shown in the bottom panel.  The
  green squares are the \griz\ photometry of the transient
  interpolated to the date of the spectrum.  The photometry was not
  used to scale the power-law continuum in the fits, so the excellent
  agreement with the derived scale factor indicates that our
  spectrophotometry is reliable.
}
\label{galsubfig}
\end{figure}

The host galaxy is spatially resolved while the transient is not, so
seeing variations mean that the amount of galaxy light relative to the
transient in the spectroscopic slit aperture cannot be reliably
determined directly from the photometry.  Instead, we fit the
observed spectra to model the host contribution.  We initially model
each epoch of spectroscopy as a linear sum of the best-fit host model
from Section~\ref{sec:sfr} and a BB with \tbb\ equal to the value
determined above from the \galex$+$PS1 SED fits and determine the
best-fit scale factors for each component.  We previously used a
similar procedure to subtract the host galaxy of PS1-10jh from its
spectra \citep{10jh}.  We then repeat our subtraction procedure using
a power-law continuum with $\alpha$=$0.73$ and find statistically
superior fits.  At each epoch, the scaling factors for the amplitudes
of the host galaxy model are identical for the BB and power law models
to within 1--3\%, indicating that our subtraction procedure is not
sensitive to this choice for the transient SED, although the scaling
factors do change if \tbb\ or $\alpha$ are varied.  An example fit to
the first \af\ spectrum is shown in Figure~\ref{galsubfig}.  The
depths of the absorption lines and amplitudes of the continuum
undulations due to the host in the model spectrum (magenta line) match
those in the data very well.

Several features of the fitting procedure are worth noting.  The first
is that we do not include any constraint from the observed photometry
(except implicitly through our choice of the power-law index or \tbb)
due to the possibility that our absolute spectrophotometry is
unreliable because of clouds or slit losses.  However, the green
squares overplotted on  Figure~\ref{galsubfig} represent the
\griz\ photometry interpolated to 
the date of the spectrum.  The excellent agreement of the photometry
with both the amplitude and color of the scaled power law is 
apparent.  This gives us confidence that the spectrophotometry is
correct and that our procedure to scale and subtract the host is
working satisfactorily.  This can be seen in another way in
Figure~\ref{sedfig}, where we  
overplot the galaxy-subtracted spectra on Days $-$5 and $+$24 along
with the photometry.  Although we apply the indicated multiplicative
offsets for clarity, no relative normalization factors were applied
between the photometry and the spectroscopy at the same epoch.  Again,
the colors and normalization of the galaxy-subtracted spectra are in
excellent agreement with the photometry.

Second, the spectra on the first two epochs exhibit a blue excess
relative to the host plus BB model that can be seen at wavelengths
below $\sim$3200~\AA\ in Figure \ref{sedfig}.
This is not simply a consequence of using an incorrect (too low)
\tbb.  If we allow \tbb\ to vary in our fits, we can find better fits
to match the overall shape of the observed spectra if \tbb\ is in the
range 25,000--30,000~K.  However, the fits compensate for the bluer
assumed BB color by increasing the amplitude of the host contribution
to match the red flux.  After subtraction of the new best-fit host
contribution, the derived transient spectrum then has a color that is
too blue relative to the colors measured from the photometry.
Equivalently, the values for \tbb\ allowed by the optical photometry
(Figure~\ref{tempfig}) are lower than those required to remove the
blue/UV excess flux.  The simplest explanation is that the blue excess
is real and the SED of \af\ is not that of a pure single-temperature
BB.  Therefore, in the remainder of this paper we use the
statistically preferred power-law model to determine the proper
scaling factors for the host galaxy.  We do not claim that the true
SED is a power law, just that a power law is a better
approximation over the limited wavelength range of our spectra for the
purpose of scaling and removing the host contribution.  We also note
that the shape of the excess is inconsistent with Balmer continuum
emission, which would be expected to peak near 3650~\AA.

Another consideration is that we subtract the spectral-synthesis model
for the host galaxy spectrum but the real spectrum could be different.
The excellent fit of the host model in Figure~\ref{hostfig}
demonstrates that any such error is small.  We repeat the subtraction
procedure using our actual host galaxy spectrum and find no
significant difference in the subtracted spectra, including the
presence of the UV excess in the fits with a BB continuum.  However,
the subtracted spectra are noticeably noisier, especially at shorter
wavelengths, so we use the model-subtracted spectra in all subsequent
analysis. 

\begin{figure}
\epsscale{1.15}
\plotone{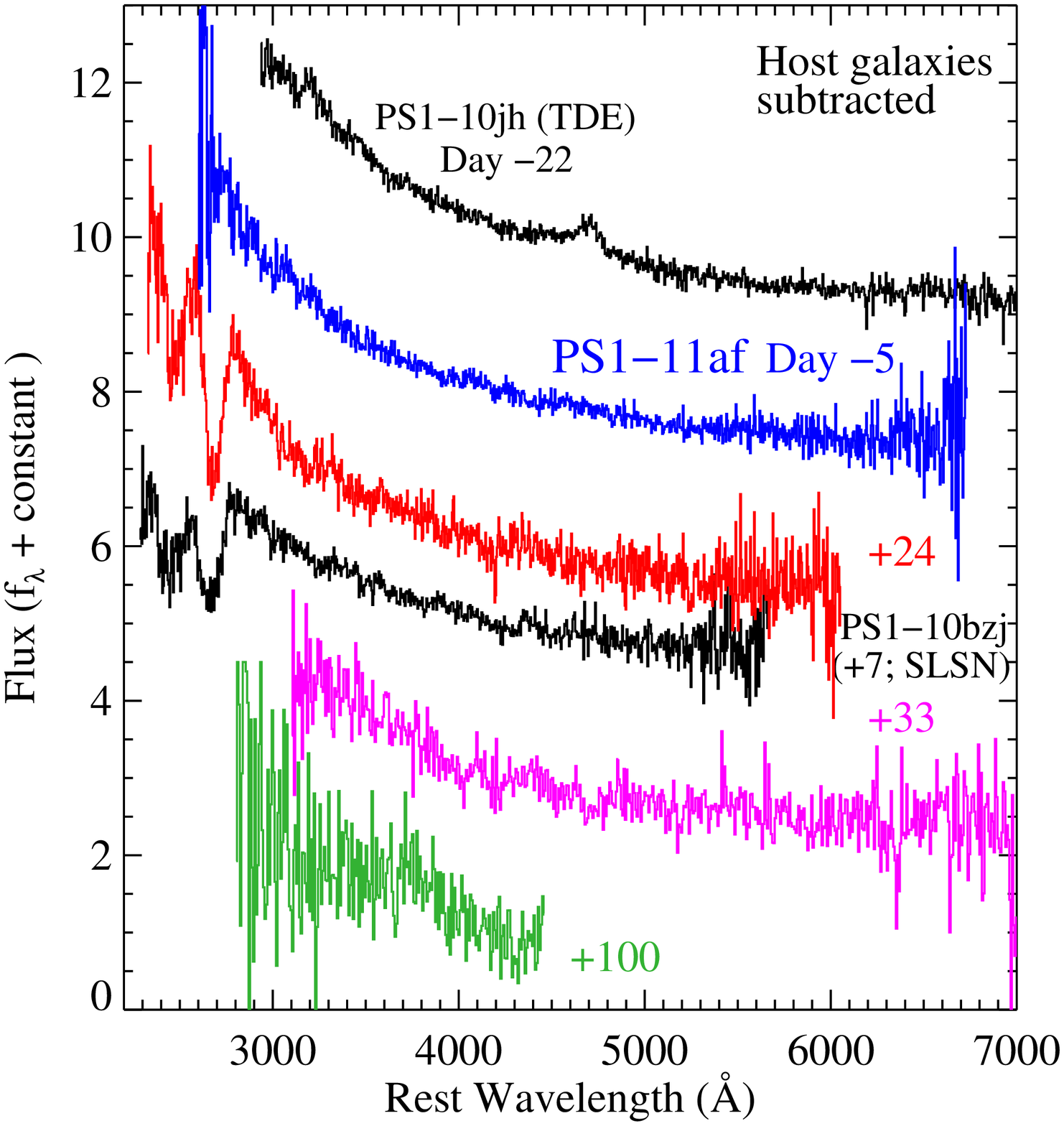}
\caption{Spectra of \af\ after subtraction of the host-galaxy
  contribution.  Note the very blue color of the spectra, with the Day
  $+$24 spectrum having deep UV absorption features.  The black
  spectrum in the middle is the $+$7~d spectrum of PS1-10bzj, which
  also shows similar features \citep{10bzj}.  The top black spectrum
  is the TDE \jh\ at Day $-$22 \citep{10jh}.  Note the broad emission
  lines from \ion{He}{2} $\lambda$4686 and $\lambda$3203 in that
  object, which are not present in any of the \af\ spectra.
}
\label{specsubfig}
\end{figure}

After determining the relative amplitude of the host galaxy
contribution for each epoch\footnote{We exclude the regions around the
  strong absorption minima on Day $+$24 from the fit.}, we subtract
the appropriately-scaled host galaxy model and present the resulting
spectral sequence in Figure~\ref{specsubfig}.  The spectra of \af\ are
very blue, with the Day $-$5 spectrum being well fit by a power law
with $\alpha$=0.75$\pm$0.02, consistent with the values measured from the
photometry, as shown in the top panel of Figure~\ref{tempfig}.  No
clear emission 
features (either narrow or broad) are present in any of our spectra,
unlike the broad \ion{He}{2} emission seen from PS1-10jh \citep{10jh}
or the broad H$\alpha$ emission detected from SDSS TDE2 \citep{vv11}.
The Day $+$24 spectrum exhibits two strong absorption features
shortward of 3000~\AA, with minima near 2450 and 2680~\AA.  The
reddest of these is clearly not present on 
Day $-$5 in our only other spectrum with overlapping wavelength
coverage.  We note that these features are definitely real (they are
apparent even in the  two-dimensional spectral frames) and are largely
unaffected by any details of the host galaxy subtraction procedure.
The host galaxy has very little contribution at these wavelengths
(e.g., Figure~\ref{galsubfig}) and the features are present prior to
host subtraction in Figure~\ref{allspecfig}.

\section{Is \af\ a SN?}
\label{sec:sn}

\af\ exhibits strong, broad UV absorption features in the Day $+$24 BC
spectrum  (Figure~\ref{specsubfig}) that are strikingly
similar to the P-Cygni absorptions present in SN atmospheres.  In
this section, we demonstrate that \af\ is unlike any known SN and
appears to be inconsistent with the expectations for any plausible SN
or explosive transient.

We start with the observation that our observed wavelength coverage
for \af\ includes the Balmer lines expected in SNe~II as well as the
diagnostic \ion{He}{1} lines of SNe~Ib ($\lambda$5876 would be
strongest), and none of these are present.  The spectra also lack the
optical features due to \ion{Fe}{2}, \ion{Ca}{2}, and the
intermediate-mass elements found in both SNe~Ia and normal SNe~Ic
\citep{fil97}.  Instead, the combination of an optical continuum
having at most weak features with strong UV absorptions is at least
qualitatively similar to the spectra of hydrogen-poor superluminous
SNe (SLSNe).  

The 
transient SCP06F6 exhibited a triplet of absorption features between
2000 and 3000~\AA\ \citep{barbary}.  The two longer-wavelength ones
were first identified by \citet{quimby11} as \ion{Si}{3} and
\ion{Mg}{2}, although spectral modeling indicates that some
\ion{Fe}{2} may contribute \citep{10bzj}.  We examine the published
spectra of SLSNe, 
and the best match we find is to the Day $+$7 spectrum of PS1-10bzj,
which we plot as a comparison in Figure~\ref{specsubfig}
\citep{10bzj}.  Most other SLSNe with optical spectroscopy at
sufficiently early epochs to match the blue colors of \af\ exhibit a
series of absorption features in the blue part of the optical that
have been identified as 
being due to \ion{O}{2}, the strongest being a ``W''-shaped absorption
near 4300~\AA\ (e.g., \citealt{quimby05ap,quimby11,10gx,laura}).  Other
objects show \ion{Fe}{2} features in the rest-frame optical
(e.g., \citealt{inserra}) and even evolve to resemble normal SNe~Ic
\citep{10gx,quimby11}.
The spectra of \af\ are instead devoid of strong features at optical
wavelengths at all available epochs.

\begin{figure}
\epsscale{1.2}
\plotone{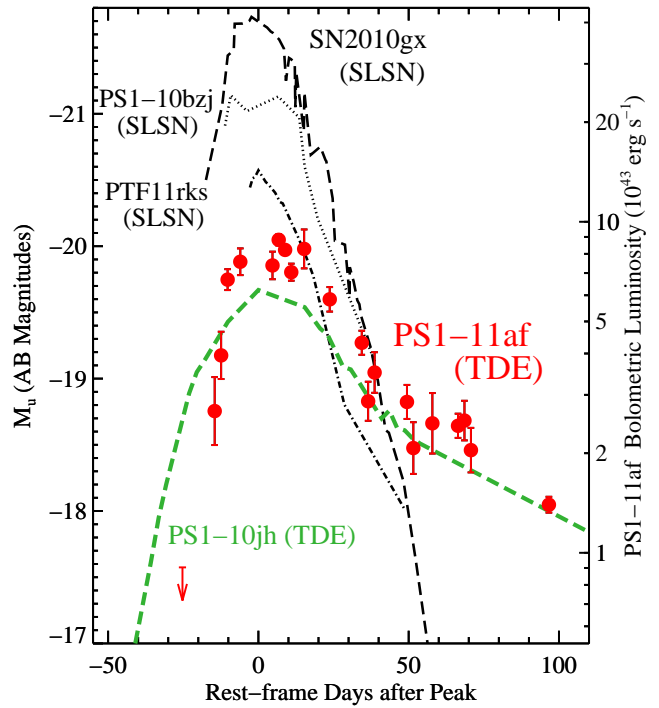}
\caption{Light curve of \af\ compared to SLSNe and \jh\ in filters
  with rest-frame wavelengths near $u$ band.  The \af\ data
  points are the \gps\ light curve
  ($\lambda_{\mathrm{rest}}$=3425~\AA) and the bolometric luminosity
  scale on the right axis represents this light curve multiplied by a
  bolometric correction factor determined from the Day $+$10 SED
  (Figure~\ref{sedfig}).  We caution that due to differing SEDs and
  bolometric correction factors, this luminosity scale only
  applies to \af.  The other objects, along with the plotted filters
  and rest-frame wavelengths, are: \jh\ (green dashed line; \gps,
  4110~\AA; \citealt{10jh}), SN 2010gx (black dashed line; $g$/$B$,
  3546/3770~\AA; \citealt{10gx,quimby11}), PS1-10bzj (black dotted
  line; \rps, 3740~\AA; \citealt{10bzj}), and PTF11rks (black
  dot-dashed line; $g$, 3900~\AA; \citealt{inserra}).
  Arrows denote 3$\sigma$ upper limits.
}
\label{bolofig}
\end{figure}

We also compare the light curve of \af\ to a few of these SLSNe in
Figure~\ref{bolofig}.  In each case, we select filters with
rest-frame wavelengths near $u$ band and correct for the distance
modulus and cosmic expansion without performing a full $K$-correction
due to the substantial uncertainties involved (e.g., PTF11rks, which
has a rather red, but poorly sampled, observed-frame $u-g$ color;
\citealt{inserra}).  The bulk of the hydrogen-poor SLSNe published to
date are significantly more luminous in $u$ than \af\ (e.g.,
\citealt{quimby05ap,quimby11,laura}).  We choose three
SLSNe at the lower end of the luminosity distribution for comparison,
but which are still more luminous in $u$ than \af.  Note that the
larger bolometric correction for the bluer SED of \af\ makes its peak
bolometric luminosity ($\sim$8$\times$10$^{43}$~erg~s$^{-1}$) closer
to that of the SLSNe ($\gtrsim$10$^{44}$~erg~s$^{-1}$) than a simple
comparison of $M_u$ implies. 

In addition, \af\ took $\sim$100~d to decline two magnitudes from peak
while the comparison SLSNe declined that much from peak in only
30--40~d.  By our \rps\ photometric point on Day $+$97, \af\ was still 
near $M_g$$\approx$$-18.25$~mag, significantly more luminous
than the objects in the sample of \citet{inserra}, with only PTF09cnd
among the published SLSNe being similarly luminous at that late epoch,
and it peaked at $M_u$$\approx$$-22$~mag \citep{quimby11}.  The faster
light curve evolution of the SLSNe is a combination of actual faster
bolometric declines along with cooling ejecta leading to a smaller
fraction of the flux being emitted at such blue wavelengths.

This implies that the  spectral correspondence may be coincidental. 
The SLSNe generally show some spectral evolution over time, so the
similarity may not necessarily hold true at other times. 
Even PS1-10bzj had a spectrum on Day $+$15 (only $\sim$8~d after the
plotted one) that had evolved in the UV and developed stronger optical
lines \citep{10bzj}, no longer appearing as similar to \af.  The
optical spectra of the SLSN~2010gx started to resemble normal SNe~Ic
by Day $+$21 \citep{10gx}, an epoch prior to that of our BC spectrum
of \af\ with the UV features.  Other, more luminous, SLSNe do evolve
more slowly spectroscopically (e.g., SCP06F6; \citealt{barbary}).
Unlike \af, none of the published SLSNe exhibiting these UV features
had a prior epoch of spectroscopy lacking them.

The spectral evolution of the SLSNe reflects the decreasing
photospheric temperature.  As mentioned above, the relatively constant
colors of \af\ are different from the rapid redward evolution of
most SNe.  The hydrogen-poor SLSNe 2010gx
and PS1-10bzj shown in Figure~\ref{colcompfig} demonstrate this
behavior.  Despite colors near maximum light that approach those
seen in \af, by $\sim$20~d after peak they are redder than \af\ is
even at Day $+$100.  The SLSN 2008es is also an interesting object for
comparison.  At early times, the spectra had a very blue continuum
with only weak \ion{He}{2} $\lambda$4686 emission \citep{gezari08es}.
Only at later times did Balmer lines from the hydrogen-rich ejecta
become apparent as the object cooled \citep{miller08es,gezari08es}.
The cooling is apparent from the redward evolution of the color curve
in Figure~\ref{colcompfig} away from the colors of \af. 

This distinction points to a fundamental difference between
\af\ and the SLSNe, so we discuss it in more detail.  In the absence
of other sources of energy input, adiabatic expansion of SN ejecta
causes them to cool.  This is especially true after maximum light,
because the photon diffusion timescale becomes shorter than the
expansion timescale of the ejecta \citep{arnett82}, and thus SNe
powered by energy sources that decline with time (such as radioactive
$^{56}$Ni or the spindown energy of magnetars) also cool after maximum
light.  Even SNe~IIP, which have additional energy input from hydrogen
recombination, cool and become redder on the plateau (e.g.,
\citealt{99em}). 

A possible exception to
these trends is provided by SNe~IIn, where ongoing circumstellar
(CSM) interaction supplies the additional energy necessary to slow the
color evolution.  There are few published UV light curves of SNe~IIn,
but we show the {\it Swift} $u-v$ color curve of SN~2007pk in
Figure~\ref{colcompfig}  \citep{07pk}. That object was initially very
blue, but rapidly became redder as the level of CSM interaction was
insufficient to prevent the ejecta shell from cooling.  A subset of
SNe~IIn exhibit much slower evolution, with long-lived CSM interaction
causing the light curve to fade very slowly.  We show the $U-V$ light
curve of a prominent recent example of such an object, SN~2010jl, in
Figure~\ref{colcompfig} \citep{zhang10jl}.  Although the color evolves
very slowly, it is $>$1~mag redder than \af.  

In addition, CSM
interaction that is sufficiently strong and long-lived to slow the
light curve evolution should also be apparent from the spectra.  The
same CSM that provides the ongoing interaction luminosity in SNe~IIn
is ionized by the SN shock prior to the interaction and produces
strong Balmer line emission \citep{chugai01}, but such lines are not
present in \af\ at any time.  Some models for SLSNe can fit their
light curves with shock breakout through a thick CSM shell, but those
models predict fast declines after the peak if there is no additional
CSM material outside of the initial shell.  If the wind continues
outside the initial shell, as is necessary for a slow decline rate, SN
IIn-like emission lines would be expected \citep{ci11,gb12}. 

The lack of color evolution of \af\ also has implications for the
interpretation of the SED that are inconsistent with the SN
hypothesis.  It is common in the literature to fit SN SEDs with a
single-temperature BB and compute \tbb\ and the implied radius of a
spherical emitting region (\rbb).  Although SN photospheres recede
through the ejecta in a comoving sense, the photospheric radii
are physically expanding, so \rbb\ generally increases with time
through maximum light and well beyond.  The luminosity still
decreases after the peak of the light curve because the cooling
\tbb\ sufficiently compensates for the increase in \rbb.  \rbb\ only
starts to decrease once the SN ejecta start to become optically thin
and the SN begins the transition to the nebular
phase. This is true for objects ranging from normal SNe II
(e.g., \citealt{99em}) and Ib/c \citep{sod08d,modjaz08d} to the SN IIn
2007pk \citep{07pk} and even SLSNe
\citep{miller08es,gezari08es,quimby11,laura,10bzj,inserra}. 

However, with \lbb$\propto$$\rbb^2$$\tbb^4$, the approximately
constant \tbb\ of \af\ implies that \rbb\ rises and falls with
\lbb\ in a manner not seen in SNe, as shown in the
bottom panel of Figure~\ref{tempfig}.  In addition, the significantly
higher \tbb\ of \af\ corresponds to significantly smaller \rbb\ at
late times.  \rbb\ for \af\ never became larger than
$\sim$1.2$\times$10$^{15}$~cm, and fell to
$\sim$6$\times$10$^{14}$~cm on Day $+$60.  By contrast, the SLSNe have
\rbb\ that grow with time to values an order of magnitude larger,
approaching 10$^{16}$~cm by $\sim$60~d after maximum light
\citep{quimby11,laura,10bzj}.  A SN 
interpretation of \af\ would have to explain not only why
\rbb\ remains so small for a source with \lbb\ comparable to SLSNe,
but how \rbb\ can fall after maximum light without \tbb\ cooling
significantly or \af\ exhibiting any signs in the SED or spectra of
the ejecta becoming optically thin in the transition to the nebular
phase.

The compactness of \rbb\ for \af\ has other implications as well.  The
P-Cygni features of SNe form in the region exterior to the
optically-thick photosphere.  In homologous expansion, the velocities
of absorption features should agree with \rbb\ of the
photosphere divided by the time since explosion, to within factors
of order unity due to radiative transfer effects (e.g.,
\citealt{kk74}).  In addition, \rbb\ for SLSNe have been shown to
increase with time at a rate approximately equal to the velocities
measured from the absorption lines \citep{quimby11,laura,10bzj}.  

At the epoch of our Day $+$24 spectrum, we estimate that \rbb\ for
\af\ was 9.2$\times$10$^{14}$~cm.  This occurred 38.5~d after the
first detection, which sets a lower limit on the time since explosion
in a SN interpretation.  In turn, this would imply that the material
at the photosphere could be moving at a maximum of 2800 \kms.
However, the UV absorption features have FWHMs of $\sim$10$^4$~\kms,
regardless of their identification.  If we identify the 2680~\AA\
absorption as being due to \ion{Mg}{2} $\lambda$2800 
doublet, the minimum of the absorption is blueshifted by
$\sim$13,000~\kms.  In other words, material moving in homologous
expansion at the velocities implied by the absorption features would
be far away (radii $\gtrsim$4.3$\times$10$^{15}$~cm) from the inferred
BB photosphere. 

Finally, the host environment of \af\ is unlike that of any known
hydrogen-poor SLSN.  The host galaxies of all hydrogen-poor SLSNe
studied to date have blue colors, strong emission lines, and other
evidence of vigorous star formation activity (e.g.,
\citealt{neill11,quimby11,laura,stoll,10bzj}).  Although we cannot exclude
the possibility of an undetectably small amount of star formation in
the local environment of \af\ around the host nucleus, the dominant
stellar population is clearly older than a Gyr
(Section~\ref{sec:sfr}).

In summary, SNe exhibit consistency between the various BB parameters
and their time evolution with their spectroscopic features, due to the
basic properties of expanding ejecta.  However, the long-lasting blue
colors of \af\ are evidence of a different time evolution driven by
different underlying physics.  The spectroscopic absorption features
are not easy to accommodate in a model with approximately thermal
ejecta in simple homologous expansion.  The fits imply that \rbb\ does
not expand and \tbb\ does not cool in the manner expected for SN
ejecta.  A SN interpretation for \af\ would have to explain the
apparently unique relationships between these
observables for this object as well as its location at the center of a
massive galaxy exhibiting no signs of recent star formation.

\section{\af\ as a TDE}

Having ruled out SNe and AGN activity as plausible explanations for
\af, we now interpret the observations in terms of a TDE.  
We estimate the expected black hole mass (\mbh) of the host
of \af\ using scaling relations from the local universe.  There is
some ambiguity in the proper 
estimate for the bulge mass because we do not fully resolve potential
substructure in the galaxy.  If we identify the central
7$\times$10$^9$~\msun\ core of the light profile with the bulge, the
relations of \citet{haring} predict
\mbh$\approx$(8$\pm$2)$\times$10$^6$~\msun, while using the total
stellar mass of the galaxy likely sets an upper limit on \mbh\ of
(1.6$\pm$0.4)$\times$10$^7$~\msun.  In the following discussion, we
initially adopt \mbh=10$^7$~\msun, with a factor of two uncertainty. 

\subsection{Optical Comparison to Previous TDEs}

\jh\ was a well-observed TDE discovered by PS1 \citep{10jh}.  It had a
similar light curve shape to \af\ (Figure~\ref{bolofig}), with
similarly long-lived blue optical colors (Figure~\ref{colcompfig}).
However, the NUV to optical color was bluer for \jh, implying a
minimum \tbb\ of 30,000~K.  \citet{10jh} argue that the
intrinsic \tbb\ had to be even higher, with a minimum of
$\sim$50,000~K being necessary to supply sufficient ionizing photons
for the observed \ion{He}{2} emission lines at early times.  A modest
amount of reddening (\ebv=0.08~mag) could reconcile 
this with the data.  

We investigate the possibility of extinction for \af\ and find that
\ebv$\approx$0.2~mag is necessary for the best-fit \tbb\ of the Day
$+$10 \galex+PS1 SED to equal 30,000~K, and doubling that value pushes
\tbb\ up to nearly 10$^5$~K.  However, such a 
large extinction would imply that \lbb\ is 1.6$\times$10$^{46}$
erg~s$^{-1}$, more than ten times the Eddington luminosity (\ledd) for
a 10$^7$ \msun\ black hole.  If we assume that the maximum permitted
\lbb\ is $\sim$3\ledd\ for \mbh$\approx$10$^7$~\msun\ (the factor of
three is to be conservative and allow for some uncertainty in the
parameters), then we can set an upper limit on the reddening of
\ebv$\lesssim$0.35~mag, which corresponds to an upper limit of
\tbb$\lesssim$5.4$\times$10$^4$~K.
While the best fit for the host galaxy SED has zero reddening
(Section~\ref{sec:sfr}), we cannot exclude the possibility of some gas
and dust local to the environment of \af\ in the nucleus.  The
broadband SED fits for \swone\ implied substantially more extinction
for the transient than was derived for the host galaxy
\citep{bloom,levan11,burrows,baz11}.  We note 
that none of our galaxy-subtracted spectra exhibit any narrow
absorption lines from intervening gas (e.g., \ion{Mg}{2}
$\lambda$2800, \ion{Ca}{2} H+K, or \ion{Na}{1} D), which argues
against a large gas column along the line of sight.

The two optically-selected SDSS TDEs of \citet{vv11} were
discovered on the decline, so there are only lower limits on the peak
luminosities, but \ttwo\ appears to be most analogous to \af.  It
peaked at $L_g$$>$4.1$\times$10$^{43}$~erg~s$^{-1}$ and had an average
\tbb=18,200~K, both of which are fairly similar to \af.  \citet{vv11}
identified several distinctive characteristics of their two events
compared to other transients and variable AGN in Stripe 82, including
the extremely blue color and slow luminosity and color evolution.
They parameterized the slow evolution using the somewhat unusual units
of $d\ln L_g/dt$ and $d\ln\tbb/dt$.  We estimate $L_g$ from $\lambda
f_{\lambda}$ in \rps\ and \tbb\ from the BB fits to the PS1
\griz\ photometry above.  We fit linear relationships to the
logarithms of both of these quantities after maximum light and find
$d\ln  L_g/dt$=($-1.8$$\pm$0.2)$\times$10$^{-2}$~d$^{-1}$ and 
$d\ln\tbb/dt$=($-$0.2$\pm$2)$\times$10$^{-3}$~d$^{-1}$, very
similar to the values for both of the SDSS TDEs.

\citet{vv11} also noted the unusual combination of the very blue color
and its slow evolution for their objects.  Our photometry does not
extend sufficiently blue to cover the SDSS $u'$ band so we cannot
directly compare our measurements to the observer-frame quantities
used by  \citet{vv11}. However, the higher redshift of \af\ allows us
to compensate somewhat for this.  We de-redshift our Day $-$5 and
$+$24 spectra to $z$=0.2 (similar to the SDSS TDEs) and use
\texttt{STSDAS/SYNPHOT}\footnote{\texttt{http://www.stsci.edu/institute/software\_hardware/stsdas/synphot}}
in IRAF to synthesize observer-frame colors.  The average colors of
\af\ from the two shifted spectra are $u'-g'$=$-0.24$~mag and
$g'-r'$=$-0.18$~mag. 
We also fit a line to the \gz\ points (fairly close to $u'-r'$ in the
rest frame) from Figure~\ref{colcompfig}
and find a slope of (0.1$\pm$3)$\times$10$^{-3}$~mag~d$^{-1}$,
consistent with no evolution.  The combination of these colors and
their stability over a long time baseline places \af\ in
the same parts of parameter space as both of the SDSS~TDEs in the
diagrams of \citet{vv11}, and away from the SNe and AGN-like
variables. This is further evidence that their objects are of a
similar class as \af. 

\citet{iya} described the discovery of PTF~10iya, a fast evolving
and UV bright nuclear flare. They modeled it
as being the result of the early super-Eddington phase of accretion
following the tidal disruption of a solar-type star by a
$\sim$10$^7$~\msun\ black hole.  That object was very different from
\af, despite a similar \tbb.  It had a somewhat higher peak luminosity
($\gtrsim$10$^{44}$~erg~s$^{-1}$, depending on the extinction
correction), but declined very rapidly ($\sim$0.3~mag~d$^{-1}$)
compared to \af.  PTF~10iya also had a bright associated 
X-ray source, with $L_X$$\approx$10$^{44}$~erg~s$^{-1}$.
Unfortunately, we have no constraints on the high-energy emission from
\af. 

The two relativistic TDEs, \swone\ and \swtwo, have much more limited
optical data (apparently because of high extinction in the case of
\swone: \citealt{bloom,baz11}).  \swtwo\ did have a slowly
evolving UV/optical light curve, but \tbb\ had a lower limit of
6$\times$10$^4$~K and \lbb$\gtrsim$10$^{45}$~erg~s$^{-1}$, indicating
a significantly more luminous accretion event than for
\af\ with a bluer SED \citep{swift2}. 

Several UV-selected TDE candidates have been found in
\galex\ observations \citep{gezari08,gezari09}.  These events had
bluer SEDs than \af\ ($\alpha$$\approx$1.1--1.4) and their inferred
\tbb\ are correspondingly hotter.  Single-temperature BB fits found
\tbb\ in the range (4.4--12)$\times$10$^4$~K \citep{gezari09}, which
is more comparable to the $\gtrsim$10$^5$~K expectations for accretion
disks near the tidal radius.  
The light curves for these objects are also generally consistent with
a $t^{-5/3}$ decline, although the lack of data points on the rise
allows for some freedom in the light curve fits.

The TDE candidate D3-13 \citep{gezari08} may be instructive for
\af.  A {\it Chandra} detection at late times required
\tbb$>$1.2$\times$10$^5$~K, while the UV/optical data at earlier times
were better fit by a separate \tbb$\sim$10$^4$~K component in a
two-temperature fit.  The 
data were not taken simultaneously, but may point to the existence of
a hotter component that is not obvious in the optical data, even if it
dominates the bolometric luminosity \citep{gezari08}.

\subsection{Light Curve Fits}

We use the bolometric light curve from Figure~\ref{bolofig} (the
observed \gps\ light curve multiplied by a bolometric correction 
to match the \lbb\ from the \galex+PS1 SED fit on Day $+$10) to
constrain the accretion properties of \af.  Third-order polynomial
fits to the data around peak give a maximum \lbol\ of
(8.5$\pm$0.2)$\times$10$^{43}$~erg~s$^{-1}$, where the errors are
derived by Monte Carlo resamplings of the observed light curve.  This
translates to \lbol/\ledd$\approx$$0.07_{-0.03}^{+0.13}$. Similarly,
we perform a trapezoidal integration of the light curve to find the
minimum total radiated energy of (4.1$\pm$0.1)$\times$10$^{50}$~erg
between the first and last detections.  If we assume a radiative
efficiency factor, $\eta$$\approx$0.1, then the minimum necessary
accreted mass onto the black hole is
$M$$\approx$0.002~(0.1/$\eta$)~\msun.  This is only a lower
limit because it does not include the tail of the light curve at late
times and, as we saw above, a small amount of extinction can
dramatically increase the flux in the UV where the peak of the SED
is and we have few constraints on the SED shape.  Still, the low value
for \lbol\ (and hence the accreted mass) is suggestive of a scenario
involving only a partial disruption of a star. 

The observed range of \rbb\ is $\sim$(5--12)$\times$10$^{14}$~cm
(Figure~\ref{tempfig}).  These values are not straightforward to
interpret in the TDE case because they were derived from the
normalization of the BB fits by assuming a spherical emitting
surface. 
The geometry of the TDE accretion flow could be quite different, with
most models assuming that a disk dominates the UV/optical continuum
unless reprocessing is invoked \citep{lu97,lr11,strubbe11,james10jh}.
Nevertheless, these 
radii provide a useful scale for the system, as they correspond to
135--400 Schwarzschild radii.  Even allowing for uncertainty in \mbh,
they are far outside the expected tidal radius for a main sequence
star.  These radii are near the tidal radius for red giant stars, but
TDEs with disruption radii that far from the central black hole are
expected to have light curves with rise times of order a year,
rather than a few weeks \citep{macleod12}.

We emphasize that these numbers have substantial systematic
uncertainties.  Most of the energy is emitted in the UV, where we only
have the limited \galex\ observations to constrain the SED and light
curve, although \tbb\ determined from the PS1 observations alone is
consistent with the fit to \galex+PS1.  
Our assumption that the SED can be approximated as a
single-temperature BB is definitely an oversimplification, as
demonstrated by the UV excess of the 
galaxy-subtracted spectra relative to a BB (Figure~\ref{sedfig}).  We
know that some UV absorption features are present on Day $+$24, near
the time of the \galex\ observations, so if any others were present in
the NUV bandpass, the UV flux might be suppressed and \tbb\ would be
underestimated. 

We have no information about
any possible emission extending to higher energies.  In the cases of
PTF~10iya and D3-13, the X-ray emission carried a comparable
or greater amount of energy to the UV/optical component without lying
on an extrapolation of the low-energy SED \citep{iya,gezari08}. If
such emission were present in \af, it would be undetectable in our
dataset.  Our derived values for \mdot\ are therefore best regarded as
lower limits.

We now consider the shapes of the light curves to see if they are
consistent with expectations from numerical modeling of stellar
disruptions.  We start with 
the estimated bolometric light curve from above and convert it to a
mass accretion rate, \mdot, assuming that $\eta$=0.1. The output
\mdot\ of numerical simulations are self similar, 
with the time and accretion rate variables being rescalable functions
of \mbh\ and the mass and radius of the disrupted star, \mstar\ and
$R_{\star}$. 

\begin{figure}
\epsscale{1.2}
\plotone{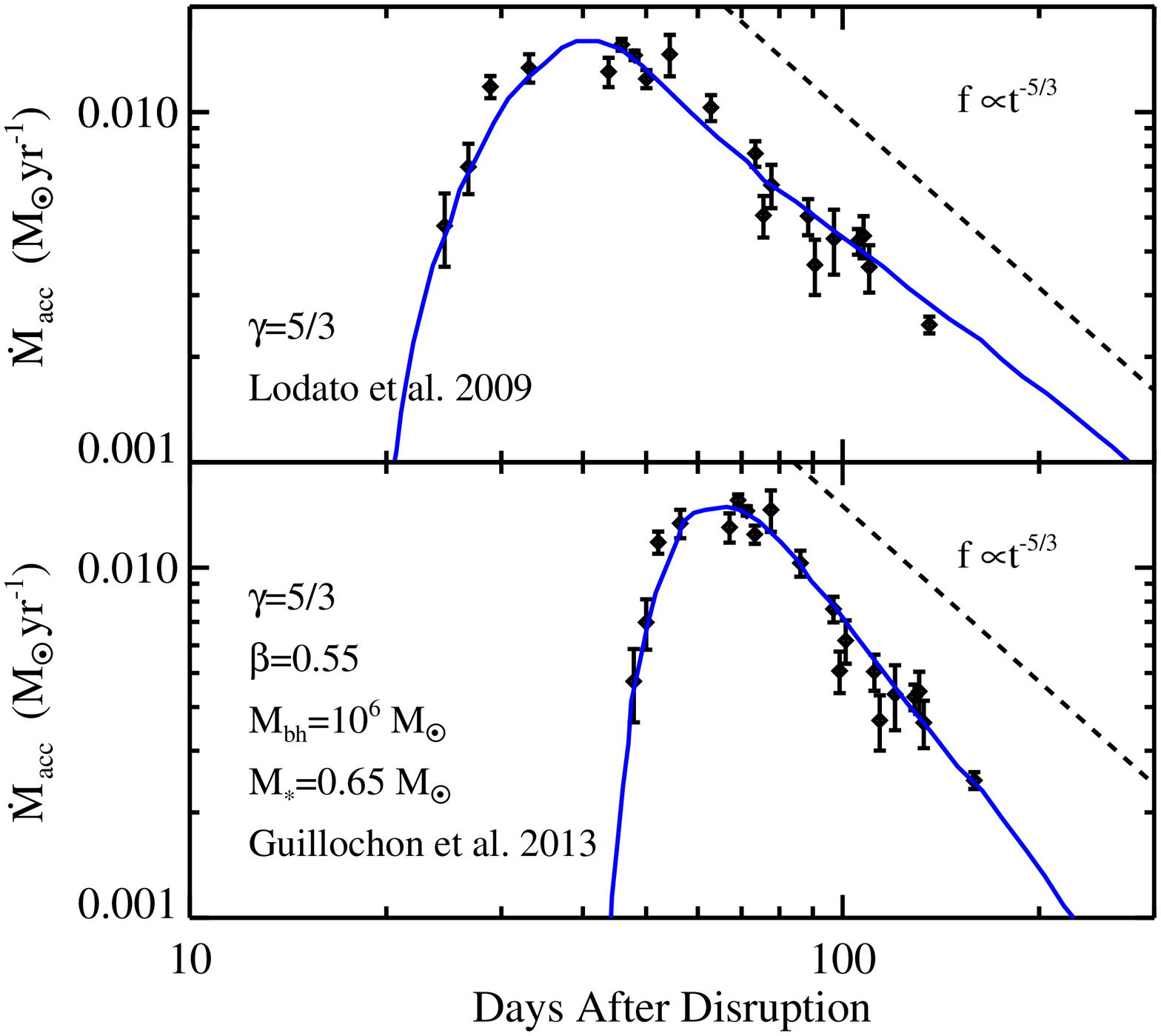}
\caption{Fits of TDE models to \mdot. We convert the bolometric light
to \mdot\ by assuming a fiducial radiative efficiency factor of
$\eta$=0.1.  The top panel shows a fit using the $\gamma$=$5/3$ model
of \citet{lodato09}, with the time of disruption occurring 39~d before
the peak of the light curve.  The bottom panel uses the partial disruption
model of \citet{grr13}, with disruption occurring 62~d before peak.
In each panel, the dashed line marks the 
traditional $t^{-5/3}$ power law expectation for TDE light curves.
}
\label{lcfit}
\end{figure}

We first fit to the simulations of \citet{lodato09}, which were
performed for the full disruption of a star 
at the tidal radius with a range of polytrope indices ($\gamma$;
assuming an equation of state with $P$$\propto$$\rho^{\gamma}$).  We
focus on models with $\gamma$=$5/3$, appropriate for low-mass stars,
because of the small accreted mass for \af.  We find an excellent
fit (top panel of Figure~\ref{lcfit}) for a disruption occurring 39~d
prior to the peak of the light curve if we scale the time variable
from the fiducial value by 0.73$\pm$0.03.  With the scalings from
\citet{lodato09}, this implies \mbh=(5.4$\pm$0.5)$\times$10$^5$
(\mstar/\msun)$^2$ ($R_{\star}$/$R_{\odot}$)$^{-3}$~\msun.  Even
for a representative low-mass main sequence star with
\mstar=0.3~\msun\ (and $R_{\star}$=0.3~$R_{\odot}$; \citealt{tout}),
\mbh\ is still $\sim$1.8$\times$10$^6$~\msun, well below our
expectations from the host scaling arguments.
Also, we had to apply a large vertical scaling factor to the plotted
curve because \mdot\ is so low.  The integral under the
curve implies a total accreted mass of only 0.003~\msun.  This is
inconsistent with the full disruption of even a 
low-mass star, which was one of the assumptions of the
\citet{lodato09} simulations. 

Therefore, we also examine partial
disruption models \citep{grr13}, which parameterize the ratio of
the tidal radius to the pericenter distance where the (partial)
disruption occurs as $\beta$.  \citet{grr13}
give scaling relations for their families of curves in terms of
several observables, but two of the parameters, the time from
disruption to the peak and the asymptotic power-law index, depend on
knowing the unobserved time of disruption.  We estimate the mass lost
by the star as 2$\times$$M_{\mathrm{acc}}$$\approx$0.006~\msun, from
which we estimate $\beta$$\approx$0.57.  We fit our light curve with
the $\beta$=0.55 model of \citet{grr13} in the bottom panel of
Figure~\ref{lcfit}, and can find good fits with \mstar=0.65~\msun\ and
\mbh=10$^6$~\msun.  This is not intended to be a complete search of
all parameter space, but instead to demonstrate that the shape of the
light curve is compatible with the curve of \mdot($t$) from a
partial-disruption event.

The fits with both sets of models prefer smaller values for \mbh\
($\sim$10$^6$~\msun) by about an order of magnitude than we expect
based on the mass of the host galaxy.  In each case, it is
driven by the short rise time of the light curve because the time axis
scales as \mbh$^{1/2}$.  Our first detection of \af\ is on Day $-$14.5
at a flux level that is a factor of $\sim$3--4 below the peak.  By
contrast, \jh\ had a rise time after the first detection of $\sim$50~d
(Figure~\ref{bolofig}; \citealt{10jh}).  This is not just a
consequence of \jh\ being brighter and more easily detectable at
fainter flux levels because it still took more than 30~d to rise the
final two magnitudes to maximum light.  \citet{10jh} found that the
same $\gamma$=$5/3$ model of \citet{lodato09} required a time stretch
factor of 1.38, compared to 0.73 for \af.  Similarly,
\citet{james10jh} found a good fit to \jh\ with their partial
disruption models with $\beta$=0.87 and \mbh=10$^7$~\msun. Their
derived \mbh\ was somewhat higher than expected from the host stellar
mass relationship, which they attribute to scatter in the
\mbh-$M_{\mathrm{bulge}}$ relationship.

A major problem with the interpretation of both fits to the \af\ light
curve is that the connection between $\dot{M}$ of stellar debris
returning to pericenter and 
the light curve in any given observed band depends on the details of
the hydrodynamics of the gas (e.g., whether an outflow forms) and the
radiative transfer, along with their evolution through the event.  We
use the observed constancy of the colors to estimate \mdot\ from the
\gps\ light curve with a constant scaling factor, but it is not clear
that this is always justified, and is certainly not expected in basic
models. 

\citet{lr11} computed multiband light curves for the \citet{lodato09}
models, including the effects of a wind \citep{strubbe09}.  None of
their models exceed $\nu L_{\nu}$ of $\sim$few$\times$10$^{42}$
erg~s$^{-1}$ in the optical band, an order of magnitude below that
observed for \af, and most are closer to 10$^{41}$~erg~s$^{-1}$.  This
is a consequence of the models having hotter spectra than \af\ and
emitting more of their radiation at higher energy.  In addition, the
model spectra evolve strongly with time, leading to light curve shapes
in each band that are quite different from the mass return rates
derived from the properties of the disrupted stars.
\citet{10jh} also found that \jh\ had a light curve shape that closely
matched the shape of the disruption models of \citet{lodato09} and
lacked the expected color evolution.  

\citet{james10jh} were able to fit the shapes of the light 
curves of \jh\ by including a reprocessing component to convert the
accretion disk luminosity to a softer component with a roughly
constant temperature, although their model invoked an unusually gray
dust extinction law.  They emphasize that pure accretion disk models 
cannot simultaneously satisfy the condition that the luminosity
directly follows \mdot\ and maintain a constant color.  The origin of
the reprocessing material is not understood.  It could result
from shocks at the disruption radius as the returning stellar debris
interacts with itself or it could represent a version of the accretion
disk winds seen in regular AGN \citep{murray95}.  Reprocessing of
some form has long been invoked to explain AGN SEDs, which also
exhibit low disk temperatures relative to the expectations of naive
thin disk models \citep{kb99}. See \citet{law12} for a recent review
of this issue and some possible solutions. 

\citet{strubbe11} predict the existence
of a TDE outflow, but only as long as the accretion rate is
super-Eddington.  After maximum light, as the rate of mass return to
pericenter drops, the optical depth in the wind drops and leaves the
hotter disk more exposed \citep{lr11}.  \tbb\ for \af\ does not evolve
in this fashion.  For our inferred accretion rate for \af\ to approach 
Eddington would require invoking some UV extinction, an unobserved
high-energy emission component (or at least more emission at shorter
wavelengths than implied by the Wien tail of our single-temperature BB
fits), or a smaller than expected \mbh.
In an outflow scenario, the slight decrease in \rbb\ after
maximum light  could be explained by the decrease in \mdot\ (and hence
\lbb) providing less radiation pressure and a weaker wind. 
\citet{strubbe11} also include the effects of the unbound material
from the disrupted star in their models, but the simulations of
\citet{james10jh} indicate that self-gravity confines the unbound
material and it has little effect on the SED or spectrum.

In summary, the shape of the light curve of \af\ can be acceptably
fit by both full and partial tidal disruption models
(Figure~\ref{lcfit}). 
However, the normalization to the low observed
luminosity requires an accreted mass that is too low for a full
disruption of a star.  This inconsistency with the full disruption
model leads us to favor a partial disruption scenario.  However, if
the bulk of the luminosity is emitted at higher energy, then the total
accreted mass could be significantly higher.  We note that the two
models shown in Figure~\ref{lcfit} have late-time decay rates that are
in one case slower and the other faster than the canonical $t^{-5/3}$
value.  This demonstrates the perils of attempting to match the
observed late-time decay rate to theoretical expectations when the
time of disruption is not observed, even for objects with light curves
that are well sampled on the rise to maximum light.

\subsection{Lack of a Relativistic Jet}

\swone\ and \swtwo\ were both discovered due to a high-energy trigger
and their long-lived, luminous X-ray counterparts were interpreted as
the results of on-axis relativistic jets
\citep{bloom,levan11,burrows,baz11,swift2}.  We have no X-ray
observations of \af, so we cannot exclude the possibility of such a
high-energy counterpart here.  However, any relativistic outflow
should produce detectable radio emission, even if it is not oriented
along our line of sight \citep{gm11}.  Most radio observations to date
of other TDEs and TDE candidates have not detected any emission
\citep{bower,vv13}.  However, there are one or two X-ray selected TDE
candidates with late-time radio emission, potentially from off-axis
jets \citep{bower}. 

\begin{figure}
\epsscale{1.2}
\plotone{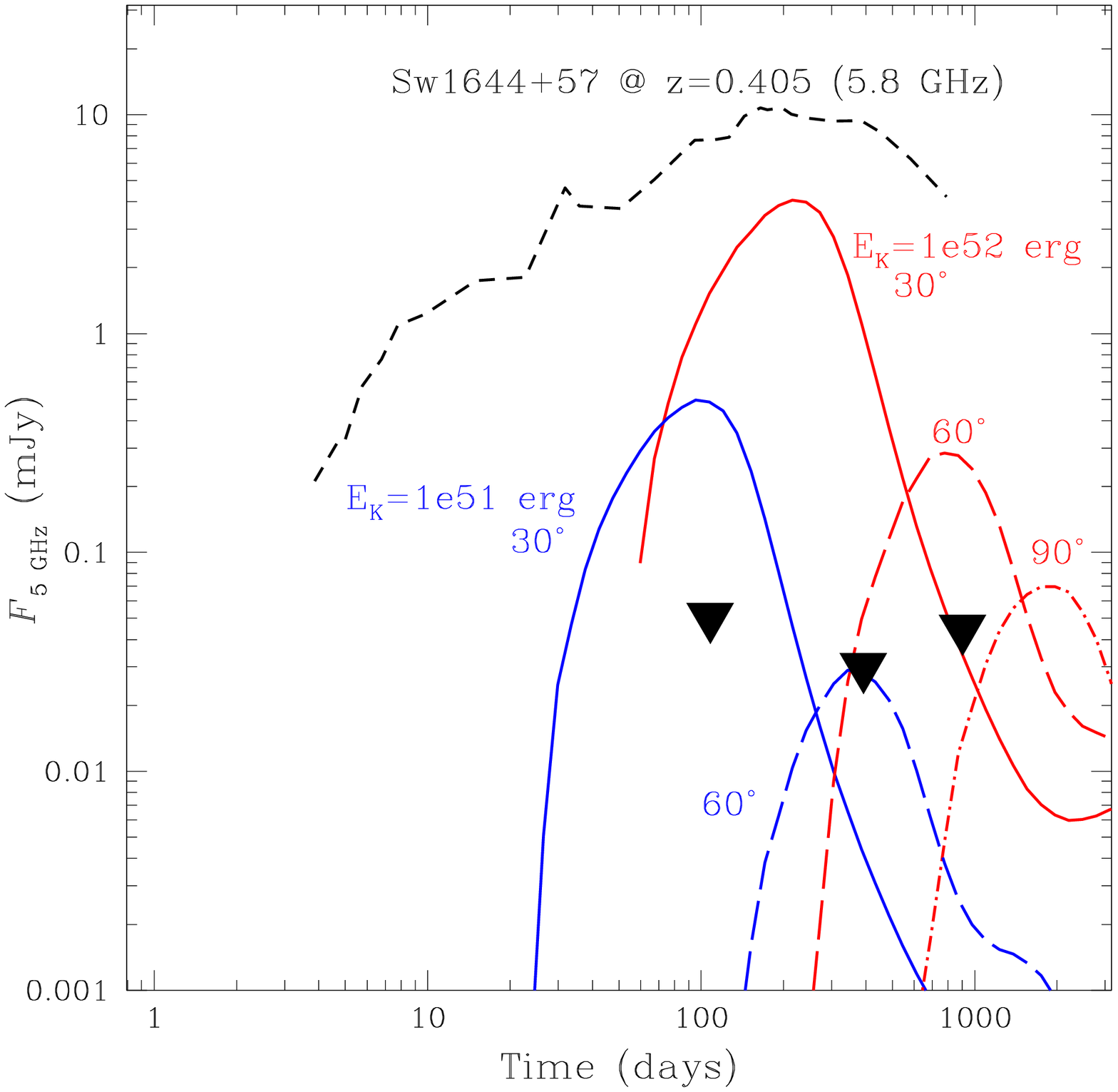}
\caption{5 GHz upper limits for \af\ (black triangles).  The abcissa
  represents observer-frame time since disruption, assumed to be 40~d
  before the optical peak for \af.  The
  dashed line is the 5.8~GHz light curve of \swone\ scaled to
  the redshift of \af\ \citep{baz11,edo12,baz13}.  The other lines
  represent impulsive GRB-like afterglow models from the BOXFIT code
  \citep{ve12}, expanding into a constant density medium
  ($n$$=$$1$~cm$^{-3}$) with an opening angle of
  $\theta_{\mathrm{jet}}$=0.1.  
  The solid and dashed lines in red have 
  $E_{\mathrm{K}}$=10$^{52}$~erg (as in \swone; \citealt{baz13}), viewed
  30$\degr$ and 60$\degr$ off-axis, respectively. The corresponding
  lines in blue have the kinetic energy scaled down by a factor of 10.
  We also show a model (red dot-dashed line) with
  $E_{\mathrm{K}}$=10$^{52}$~erg viewed 90$\degr$ off-axis (actually
  89.95$\degr$ for numerical reasons).
  In all models, the microphysical parameters representing the
  fractions of energy in 
  the electrons and magnetic fields, $\epsilon_{\mathrm e}$ and
  $\epsilon_{\mathrm B}$, were fixed to match the best-fit values for \swone\
  of 0.1 and 0.01, respectively \citep{baz13}.
}
\label{radfig}
\end{figure}

Our three epochs of VLA non-detections strongly constrain the presence
of any relativistic outflow. The redshift of \af\ is only slightly
higher than that of \swone, so 
an equivalently powerful relativistic jet would be easily detectable,
even if it were viewed off-axis.  
As shown in Figure~\ref{radfig}, the
peak 5.8~GHz radio flux of \swone\ \citep{baz11,edo12} is a factor of
100--300 above our non-detections of \af\ on similar timescales.

Several models for the radio emission from jets produced by TDEs exist
in the literature.  One class invokes an analogy to
gamma-ray burst (GRB) afterglows, with both reverse and forward shocks
\citep{gm11,metzger} from the decelerating blast wave potentially
contributing to the observed radio emission.  An alternative
invokes internal shocks within the jet, by analogy to AGN jets
\citep{vv11b}. 
 A relativistic outflow with a
kinetic energy of $E_{\mathrm{K}}$ becomes
non-relativistic on a timescale of
$\sim$300~($E_{\mathrm{K}}/10^{52}$~erg)$^{1/3}n^{-1/3}$~d, where $n$
is the circumnuclear 
density, assumed to be uniform (e.g., \citealt{bower}).  On that
timescale, comparable to that of our late-time observations of \af,
the emission becomes more easily detectable by off-axis observers.  

We take advantage of this fact to set limits on the presence of a
relativistic jet by generating light curves using the GRB afterglow
models produced by the BOXFIT code of \citet{ve12} rather than making
detailed TDE jet models.  The most important difference between jets 
produced by GRBs and by TDEs is that the former have a single
impulsive episode of energy injection and the latter can have energy
injection extending to late times \citep{edo12,decolle}. 
However, we assume that this distinction primarily affects the shapes
of the model light curves and is less important for predicting the
timing and flux of the peak in the radio band, which mostly reflect
the total energy in the jet and its orientation relative to our line
of sight \citep{vm12}.

We limit our models to those with microphysical parameters fixed to
the best-fit values from afterglow models for
\swone\ \citep{edo12,baz13}.  BOXFIT assumes a constant density medium
that we set to $n$$=$$1$~cm$^{-3}$.  We initially use
$E_{\mathrm{K}}$=10$^{52}$~erg to match the total energy in the jet
measured for \swone\ for an assumed opening angle of
$\theta_{\mathrm{jet}}=0.1$ \citep{baz13}. Light curves observed
30$\degr$ and 60$\degr$ from the 
axis of such a jet are shown in Figure~\ref{radfig}
and the peak fluxes clearly violate our upper limits for \af.  If we
scale the energy in the jet down by a factor of 10, the 60$\degr$
off-axis case is only marginally consistent with the data. 
Also, a jet as powerful as \swone\ oriented in the plane of the sky
(90$\degr$ off-axis) could still be consistent with our limits because
the radio peak is pushed to even later times (as expected;
\citealt{vm12}). An unusually low density medium could also suppress
the radio emission.
We defer more detailed consideration of the parameter space excluded
by our limits on off-axis jet production to future work.

Jet formation in TDEs could be strongly tied to the accretion rate
relative to Eddington \citep{gm11,decolle,sasha,vv13,bower}.  By
analogy with  X-ray binaries, \citet{sasha} hypothesize that jets are
features of either super-Eddington or strongly sub-Eddington phases.
The jet in \swone\ can be modeled by assuming that energy injection
ended when the accretion rate dropped below a threshold value
\citep{baz13,decolle}. 
One possibility for \af\ not forming a jet is that the peak accretion
rate of $\sim$0.07~$\dot{M}_{\mathrm{Edd}}$ simply never
reached a sufficiently high value.  If \mbh\ really is closer
to 10$^6$~\msun, as derived from the light curve fits, or if there is
a significant unobserved high-energy emission component, then
\mdot\ would be closer to the Eddington value and the lack of jet
formation would have to be tied to some other parameter, such as the
black hole spin or the magnetization of the stellar debris.  Only
$\sim$10\% of optically-luminous quasars are radio loud \citep{k89},
and a similar fraction could be applicable to TDEs \citep{bower}.

\subsection{Origin of Transient UV Absorption Features}
\label{sec:line}

\begin{figure}
\epsscale{1.2}
\plotone{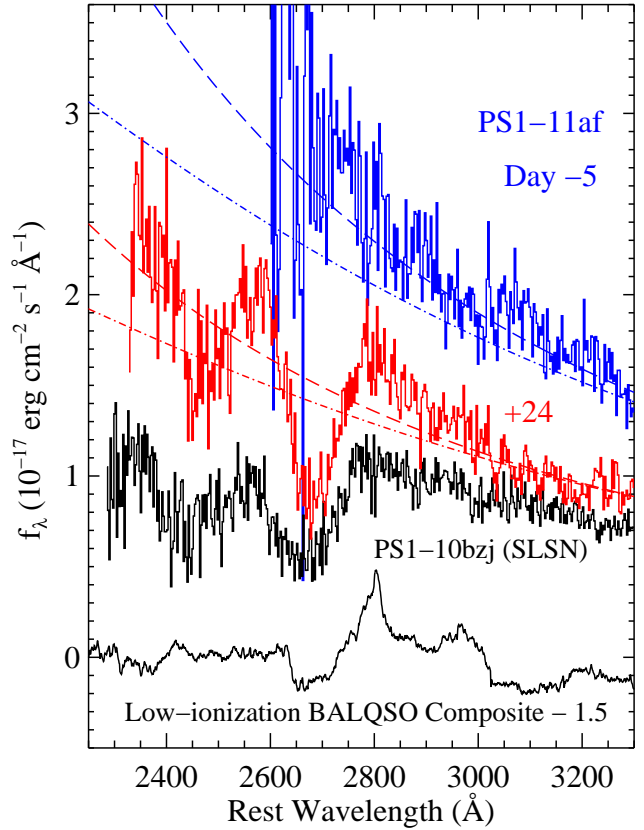}
\caption{Broad UV absorption features in the Day $+$24 spectrum of
  \af\ (red).  The Day $-$5 spectrum (blue) lacks any features.  The
  top black spectrum is the same Day $+$7 spectrum of PS1-10bzj shown
  in previous plots \citep{10bzj}.  Although multiplicative scaling
  factors have been applied for clarity, no constant offsets have been
  added to the \af\ or PS1-10bzj spectra.  Note that the PS1-10bzj
  spectrum exhibits narrow \ion{Mg}{2} absorption from gas along the
  line of sight which is not present in either of the \af\ spectra.
  The bottom black spectrum
  is the composite low-ionization BALQSO spectrum from
  \citet{brotherton}, shifted down by the indicated amount.
    The dashed lines for
  each of the \af\ spectra are power-law fits to the continuum between
  3000--6000~\AA. 
  The dot-dashed lines are \tbb=19,080~K BB spectra normalized to
  match the continuum at 3500~\AA.  See Section~\ref{sec:line} for
  discussion. 
}
\label{linefig}
\end{figure}

We now return to the most novel aspect of \af\ compared to
previously-observed TDEs and TDE candidates, the transient broad UV
absorption features.  A zoom-in on the two absorption features is
shown in Figure~\ref{linefig}. 
We fit Gaussians to the two absorption features in the
galaxy-subtracted Day $+$24 spectrum.  The shortest-wavelength one has
a centroid of 2470~\AA\ and a FWHM of 10,100$\pm$1200~\kms.  The other
one is centered at 2680~\AA\ with a FWHM of 10,200$\pm$400~\kms.  The
rest-frame equivalent widths are $\sim$25 and 50~\AA, respectively.
No additive offset has been applied to the flux scale in
Figure~\ref{linefig}, so the zeropoint is appropriate for the
spectra.  The 2680~\AA\ absorption has a minimum at about half of the
interpolated continuum flux.  Although the Day $-$5 LDSS spectrum
becomes noisy very rapidly at shorter wavelengths due to the low
instrument sensitivity in the blue, the 2680~\AA\ absorption would be
quite prominent if it were present.

The redshift of \af\ is higher than most of the other TDEs,
so the very few available spectra for other objects do not generally
cover the wavelengths of these features and we cannot determine
whether they are present.  For example, the spectra of
\jh\ and \ttwo\ do not extend sufficiently far to the UV 
(Figure~\ref{specsubfig}; \citealt{10jh,vv11}).  An important exception
is \swtwo, which is at the much higher redshift of 1.1853, and has no
obvious broad spectral features in the rest-frame UV despite 
high-quality spectra \citep{swift2}.  The 2680~\AA\
feature would be on the blue edge of the observed wavelength range
for the first spectrum of PTF~10iya, but does not appear to be present
\citep{iya}. 

\af\ also lacks the optical emission lines seen in some other
objects. \jh\ exhibited broad \ion{He}{2} $\lambda$4686 and
$\lambda$3203 
emission prior to maximum light, with FWHMs of 9,000$\pm$700~\kms
\citep{10jh}, and \ttwo\ had a broad H$\alpha$ line after maximum
light with a FWHM of 8000~\kms \citep{vv11}.  None of these lines
appear in any of our \af\ spectra at any epoch, and it is clear from
Figure~\ref{specsubfig} that we would have easily been able to detect a
\ion{He}{2} $\lambda$4686 emission line with the same equivalent width
as that seen in \jh.  \citet{10jh} argued
that the observed SED of \jh\ did not supply enough ionizing photons
to explain the observed \ion{He}{2} line fluxes, and so invoked a
modest amount of extinction and a higher intrinsic \tbb.  \af\ is not
as blue as \jh, so one possibility is simply that it has a lower
intrinsic \tbb\ and cannot ionize sufficient helium for detectable
emission.  It is also perhaps not a 
coincidence that the FWHMs of the absorption features in \af\ are
similar to the widths of the emission features in the other objects
(and in AGN) and may indicate that the line-forming regions are at
similar distances from the central black hole (in units normalized to
\mbh). 

We now consider possible identifications for these features.  The red
wing of the 2680~\AA\ absorption rejoins the apparent continuum level
near 2800~\AA\ (Figure~\ref{linefig}). The \ion{Mg}{2} $\lambda$2800
doublet is a very natural identification for this feature because it
is strong in a wide variety of astronomical objects and there are few
strong isolated lines at nearby wavelengths.  As stated in
Section~\ref{sec:sn}, the absorption minimum is blueshifted by
13,000~\kms, a reasonable value given the FWHM of the feature.  If the
other feature is blueshifted by a similar amount, it implies that the
rest wavelength of the absorbing ion should be near 2578~\AA.  One
possible candidate is \ion{Fe}{2}, which has strong resonance lines at
$\lambda$2586 and $\lambda$2600.  However, the feature at similar
wavelengths seen in SLSNe such as our PS1-10bzj comparison has been
identified as \ion{Si}{3} $\lambda$2542, following \citet{quimby11},
and supported by detailed radiative transfer models \citep{dessart12},
although some \ion{Fe}{2} may contribute to the blend \citep{10bzj}.

\citet{strubbe11} made predictions for spectral signatures of a wind
produced by super-Eddington accretion at early times in TDEs.  We
inspect their output spectra and the only line produced between 2500
and 3000~\AA\ in any of their models is \ion{Mg}{2}.  Their models
with the \ion{Mg}{2} absorption also predict stronger Balmer and
optical \ion{He}{2} absorption, which we do not observe.  Their models
assume a hot input spectrum (\tbb$\gtrsim$10$^5$~K) from the disk at
the base of the wind leading to significantly hotter output spectra 
than we measure for \af\ and hence a higher degree of ionization in
their spectra. We conclude that the \ion{Mg}{2} identification for the
2680~\AA\ absorption is robust, but regard the other one as more
uncertain. 

Broad UV absorptions are seen in $\sim$10\% of quasars, known as the 
BALQSOs \citep{weymann}.  Approximately 15\% of these exhibit
absorption in low-ionization lines, including the same \ion{Mg}{2} and
\ion{Fe}{2} lines as those (possibly) present in \af\ \citep{voit}.
We show a composite spectrum of low-ionization BALQSOs
\citep{brotherton} in Figure~\ref{linefig} as a comparison.  This may
be somewhat misleading, as the process of making a composite spectrum
averages over discrete absorptions that are frequently narrower or
detached from the rest wavelength.  The typical shapes of the
absorption troughs in BALQSOs (e.g., \citealt{voit,hall02}) are
generally quite different from the single broad absorption with a
smooth profile for each line in \af.   At lower luminosities, the
intrinsic UV absorption lines in Seyfert 1 galaxies tend to be narrow
and located at much lower velocities than the $\sim$10$^4$~\kms\ we
see here \citep{crenshaw}.

Low-ionization BALQSOs are on
average redder than normal quasars due to dust extinction in the
absorbers \citep{sf92,brotherton}, but \af\ is instead bluer 
than normal quasar SEDs.  While the high-ionization absorption lines
in BALQSOs may be formed coincident with the BLR \citep{murray95},
recent photoionization work has determined that the
low-ionization absorbers are located several kpc from the central
black hole (e.g., \citealt{moe}).  We conclude that although the
absorption  features in \af\ potentially come from similar ions as
those seen in low-ionization BALQSOs, the physical situation is very
different. 

Assuming a virial equilibrium with $v^2 \approx G\mbh/R$ and
\mbh=10$^7$~\msun, the expected typical velocities ($v$) for material 
located near our derived \rbb\ on Day $+$24 are 12,000~\kms.  This is
impressively close to the measured absorption blueshift of the
\ion{Mg}{2} line, given the uncertainties in \mbh.  There are also
geometric uncertainties in the interpretation of \rbb\ in an accretion
scenario and the factors of order unity involved 
in relating the absorption velocities to the virial velocity if they are
formed in some sort of outflow with an asymptotic velocity that is
a fraction of the escape speed.  Still, this approximate equivalence,
along with the fact that the line absorption is completely blueshifted
from the rest wavelength, is evidence that the line formation region
is in an outflow just outside of the continuum formation region.

\citet{james10jh} argued that the lack of hydrogen lines in
\jh\ can be explained by high ionization in the line-formation
region.  They analogize to reverberation mapping results from AGN
\citep{bm82,peterson04}, which imply that the typical distances of
H$\alpha$ and H$\beta$ emission from the central black hole for AGNs
with the continuum luminosity of \jh\ are larger 
than the outer radius of the debris disk from the disrupted star.
Such an explanation is harder to understand here if our line
identifications are correct.  It is true that our maximum values for
\rbb\ are 1.2$\times$10$^{15}$~cm ($\sim$0.5 light-days), which is
significantly closer than the typical H$\beta$ lags of $\sim$10~d for
AGN with continuum luminosities similar to that of
\af\ \citep{peterson04}. 
However, the \ion{Mg}{2} $\lambda$2800 emissivity
of BLR clouds closely tracks that of H$\beta$ under a wide variety of
density, ionizing flux, and ionizing SED assumptions
\citep{korista97}, and empirically, the FWHMs of the two emission
lines are identical \citep{mj02}.  This implies that the formation
regions for these two lines should be very similar.  Our spectrum with
the likely \ion{Mg}{2} absorption lacks any evidence of H$\beta$ in
emission or absorption (and does not extend sufficiently far to the
red to include H$\alpha$).  Furthermore, the \ion{Fe}{2} emission (at
least in the optical) has been shown to originate in the outer parts
of the BLR, farther than the H$\beta$ region \citep{barth13}, although
the only available reverberation lag measurement for the same resonant
UV1 \ion{Fe}{2} multiplet that we likely see here suggests that it may
form closer to the high-ionization \ion{C}{4} and Ly$\alpha$ lines
\citep{maoz93}.

One possible way to understand the simultaneous appearance of
\ion{Mg}{2} and the lack of Balmer lines starts from the observation
that our spectrum with the line features lacks strong \ion{Mg}{2} {\it
  emission} as well (cf. the BALQSO composite in
Figure~\ref{linefig}).  The balance between the emission and
absorption parts of the profile depends on the optical depths and
relative importance of scattering versus true absorption in a
particular line, as well as the geometry and density profile of the
emitting region.  In pure scattering and a spherical geometry for an
outflow, we might expect to see some SN-like P-Cygni emission as well
as absorption if there is an outflow.  

\citet{strubbe11} instead argued that the optical depths of resonance
lines in the TDE outflow will be small, and in contrast to the
line-driven disk wind models for BALQSOs (e.g., \citealt{murray95}),
the absorption region in TDE outflows will form close to the continuum
photosphere, which limits the geometric extent of the emission
region and the equivalent width of any emission.  Although the spectra
of \af\ lack a clear quasar-like \ion{Mg}{2} emission line, there is
weak evidence for some broad emission. 
In Figure~\ref{linefig}, we fit power laws to the 3000--6000~\AA\
continua and extrapolate them to the UV (dashed lines).  On Day $-$5,
the power law remains a decent fit at bluer wavelengths (with a little
excess emission). However, on Day $+$24, there appears to be some
excess relative to the power law near 2600 and 2800~\AA, to the red of
each of the absorption lines, with the continuum returning to match
the best fit power law near 2400~\AA.  This is only weak evidence
because it is sensitive to our assumption about the intrinsic
continuum shape.  If true, it is also possible that some of the UV
excess relative to a BB seen in Figure~\ref{sedfig} is actually a
superposition of numerous weak emission features, as in the ``Little
Blue Bump'' of quasars. 

Observationally, some SNe have suppressed Balmer P-Cygni features at
early times when the ejecta are hot, even when the ejecta are known to
be hydrogen rich.  The SLSN 2008es, which we used above as comparison
for the color evolution of \af, is a good example.  The Balmer lines
did not become distinct until \tbb\ dropped below $\sim$15,000~K,
below the temperatures measured for 
\af\ \citep{miller08es,gezari08es}.  Even in normal SNe II at very
early times, a similar effect is present when \tbb\ is
near 2$\times$10$^4$~K and very steep density gradients in the outer
ejecta lead to a very small line formation region relative to the
continuum photosphere \citep{dessart08}, but such an explanation would
not apply in a situation with deep absorptions, where the line
formation region is likely extended.  Nevertheless, perhaps it is
possible for radiative transfer effects to suppress the hydrogen
absorption below the expectations of \citet{strubbe11}, although more
detailed work is needed to verify this.

\section{Conclusions}

We have presented observations of the UV-bright transient \af, which
was discovered by PS1 and also detected by \galex.  The transient is
coincident with the nucleus of
a quiescent early-type galaxy with no evidence for either AGN activity
or star formation.  \af\ was detected by PS1 for almost four
months in the rest frame and had unusually blue colors the entire
time.  A BB fit to the \galex+PS1 SED gave \tbb=19,000~K, with little
sign of evolution over the course of observations.  The large
amplitude of the transient, combined with the lack of variability of
the host in other observing seasons and the very blue colors, are
inconsistent with an AGN interpretation.

Multiepoch spectroscopy of \af\ at early times revealed 
several unusual features.  At Day $-$5, the spectra were completely
featureless and were well fit by a power-law continuum with
$f_{\nu}$$\propto$$\nu^{0.75}$.  By Day $+$24, two broad UV
absorption features became apparent.  These features are strikingly
similar to the P-Cygni absorption features in the UV seen in some
classes of SLSNe.  However, the derived BB parameters for the SED of
the transient are hard to accommodate in a SN
interpretation. \rbb\ does not expand, and \tbb\ does not cool.
Moreover, the apparent velocities of the absorption features are too
high to correspond to material in homologous expansion at the
photospheric radius implied by \rbb.

The basic observables of the colors and optical luminosities, as well
as the slow evolution of both, are comparable to values reported for
the previous optically-selected TDEs, \jh\ \citep{10jh} and the two
from SDSS \citep{vv11}. The slow evolution is unlike the
fast-declining event PTF~10iya \citep{iya}. We can fit
the shape of the light curve with models for \mdot\ from TDEs
\citep{lodato09,grr13}, but we set a lower limit on the accreted mass of
$\sim$0.002~\msun, which is indicative of a partial disruption event. 
We lack any constraints on emission shortward of the \galex\ NUV
band, which could substantially raise the inferred \mdot.
Our three
epochs of non-detections from the VLA over the course of two years
after the disruption set strong constraints on the existence of any
relativistic outflow, even one that is off-axis.

 The relatively low and constant
\tbb\ measured for \af\ and \jh\ require that the majority of the
optical light 
is reprocessed from the accretion disk, which would otherwise be
much hotter and emit a spectrum that evolves with time
\citep{strubbe09,strubbe11,lr11,james10jh}.
Our observations do not constrain the structure of the reprocessing
component, but the broad blueshifted UV absorption features point to
an outflow.  Outflows have 
been predicted for the early super-Eddington phase of TDEs
\citep{strubbe09}, but our basic measured parameters imply that the
peak accretion rate for \af\ is sub-Eddington.   This contradiction
can be avoided if the majority of the luminosity is emitted at 
higher energies by another spectral component, if there is significant
extinction, or if \mbh\ is substantially lower than predicted by local
scaling relationships.  The last of these possibilities is perhaps the
most exciting, as TDEs offer the promise of being one of the few
probes of \mbh\ in distant quiescent galaxies.  
Clearly, self-consistent models for the
formation of the reprocessing component are necessary if we wish to
confidently use the observed properties of TDEs to study quiescent
black holes in distant galaxies. 

However, much
uncertainty remains in the interpretation of the TDE light curves. 
Future objects would benefit from simultaneous observations at higher
energies (such as with {\it Chandra}) to constrain the relative
contributions of the hotter emission from a disk and the reprocessing
component deduced from the cooler \tbb\ seen at UV and optical
wavelengths. 
  The connection between these cooler
optically-selected TDEs such as \af\ and previously reported soft
X-ray and UV flares with \tbb$\gtrsim$10$^5$~K is unclear (e.g.,
\citealt{komossa99,gezari09}).  

Finally, it is interesting to consider the observable properties of
\af\ at different redshifts.  At $z$$\approx$0.2, similar to
previous optically-selected TDEs, the broad UV absorption features
would not be accessible to most ground-based spectrographs.  The
transient would then have a featureless blue optical spectrum
and would be associated with the nucleus of an early-type galaxy and
thus would probably not arouse suspicions of being a SN.  Conversely,
at higher redshifts, as anticipated for objects in the LSST era,
the UV features of \af\ would be more easily observable, possibly
along with others at shorter wavelengths.  If such an event occurred in
a galaxy exhibiting nebular emission lines or other evidence of star
formation, it seems likely that the transient would be confused with
a SN.  In such a scenario, careful consideration of the SED of the
transient and its (lack of) evolution in the LSST colors would be
necessary to discriminate between the two interpretations.  At the
lower spatial resolution of observations at those higher redshifts,
the number of SNe with positions consistent with the nucleus is
already higher than the expected TDE rate \citep{strubbe11}, making
the search for TDEs similar to \af\ in the LSST dataset challenging.

\acknowledgments
We thank the staffs at PS1, Magellan, Gemini, the MMT, and the VLA for 
their assistance with scheduling and performing these observations. 
We acknowledge useful discussions with E. Ramirez-Ruiz, the assistance
of T. Laskar with some of the MMT observations, and the help of
A. Monson with FourStar data reduction. 
BAZ is supported by an NSF Astronomy and Astrophysics Postdoctoral
Fellowship under award AST-1302954. 
This paper includes data gathered with the 6.5-m Magellan Telescopes 
located at Las Campanas Observatory, Chile.  Some observations
reported here were obtained at the MMT Observatory, a joint facility
of the Smithsonian Institution and the University of Arizona. 
The Pan-STARRS1 Surveys (PS1) have been made possible through the
contributions of the Institute for Astronomy, the University of
Hawaii, the Pan-STARRS Project Office, the Max-Planck Society and its
participating institutes, the Max Planck Institute for Astronomy,
Heidelberg and the Max Planck Institute for Extraterrestrial Physics,
Garching, the Johns Hopkins University, Durham University, the
University of Edinburgh, Queen's University Belfast, the
Harvard-Smithsonian Center for Astrophysics, the Las Cumbres
Observatory Global Telescope Network Incorporated, the National
Central University of Taiwan, the Space Telescope Science Institute,
the National Aeronautics and Space Administration under Grant
No. NNX08AR22G issued through the Planetary Science Division of the
NASA Science Mission Directorate, the National Science Foundation
under Grant No. AST-1238877, and the University of Maryland. 
Some observations were obtained under Program ID GS-2011A-Q-29 (PI:
Berger) at the Gemini Observatory, which is operated by the 
    Association of Universities for Research in Astronomy, Inc., under
    a cooperative agreement  
    with the NSF on behalf of the Gemini partnership: the National
    Science Foundation  
    (United States), the National Research Council (Canada), CONICYT
    (Chile), the Australian  
    Research Council (Australia), Minist\'{e}rio da Ci\^{e}ncia,
    Tecnologia e Inova\c{c}\~{a}o  
    (Brazil) and Ministerio de Ciencia, Tecnolog\'{i}a e
    Innovaci\'{o}n Productiva (Argentina). 
SJS acknowledges funding from the European Research 
Council under the European Union's Seventh Framework Programme
(FP7/2007-2013)/ERC Grant agreement n$^{\rm o}$ 291222.
The National Radio Astronomy Observatory is a facility of the National
Science Foundation operated under cooperative agreement by Associated 
Universities, Inc.  
RPK's work on SNe is supported in part by NSF grant AST-1211196.
\texttt{STSDAS} is a product of the Space Telescope Science
Institute, which is operated by AURA for NASA. 
Development of the BOXFIT code was supported in part by NASA through
grant NNX10AF62G issued through the Astrophysics Theory Program and by
the NSF through grant AST-1009863.  
Some of the computations in this paper were run on the Odyssey cluster 
supported by the FAS Science Division Research Computing Group at
Harvard University.

{\it Facilities:} \facility{PS1 (GPC1)}, \facility{Magellan:Baade
  (IMACS, FourStar)}, \facility{Magellan:Clay (LDSS3)},
  \facility{MMT (Blue Channel spectrograph)},
  \facility{Gemini:South (GMOS-S)}, \facility{EVLA}

\LongTables

\clearpage

\begin{deluxetable}{lrcccc}
\tablecaption{\af\ Photometry}
\tablehead{
\colhead{MJD} &
\colhead{Epoch\tablenotemark{a}} &
\colhead{Filter} &
\colhead{Magnitude\tablenotemark{b}} &
\colhead{Error} & 
\colhead{Instrument} \\
 & \colhead{(d)} & & (AB) & &
}
\startdata
55592.5 &   8.2 & NUV & 21.55 & 0.13 &  \galex\ \\
55594.3 &   9.5 & NUV & 22.07 & 0.26 &  \galex\ \\
55596.8 &  11.2 & NUV & 21.59 & 0.17 &  \galex\ \\
55598.7 &  12.6 & NUV & 21.89 & 0.20 &  \galex\ \\
55600.6 &  14.0 & NUV & 21.87 & 0.17 &  \galex\ \\
55602.5 &  15.3 & NUV & 22.16 & 0.20 &  \galex\ \\
55604.4 &  16.7 & NUV & 22.04 & 0.18 &  \galex\ \\
55606.4 &  18.1 & NUV & 21.81 & 0.15 &  \galex\ \\
55608.2 &  19.4 & NUV & 22.21 & 0.22 &  \galex\ \\
\hline
55545.5 & $-$25.3 & \gps & $>$23.83 & \nodata & PS1 \\
55560.6 & $-$14.5 & \gps & 22.65 & 0.26 &  PS1 \\
55563.6 & $-$12.4 & \gps & 22.23 & 0.18 &  PS1 \\
55566.6 & $-$10.3 & \gps & 21.65 & 0.08 &  PS1 \\
55572.5 &  $-$6.1 & \gps & 21.52 & 0.10 &  PS1 \\
55587.6 &   4.7 & \gps & 21.55 & 0.10 &  PS1 \\
55590.5 &   6.8 & \gps & 21.35 & 0.04 &  PS1 \\
55593.5 &   8.9 & \gps & 21.43 & 0.03 &  PS1 \\
55596.4 &  11.0 & \gps & 21.60 & 0.06 &  PS1 \\
55602.4 &  15.2 & \gps & 21.42 & 0.15 &  PS1 \\
55614.3 &  23.7 & \gps & 21.80 & 0.09 &  PS1 \\
55629.3 &  34.4 & \gps & 22.13 & 0.09 &  PS1 \\
55632.3 &  36.5 & \gps & 22.57 & 0.15 &  PS1 \\
55635.3 &  38.7 & \gps & 22.36 & 0.15 &  PS1 \\
55650.4 &  49.4 & \gps & 22.58 & 0.13 &  PS1 \\
55653.4 &  51.5 & \gps & 22.93 & 0.20 &  PS1 \\
55662.3 &  57.9 & \gps & 22.74 & 0.23 &  PS1 \\
55674.3 &  66.4 & \gps & 22.76 & 0.09 &  PS1 \\
55677.3 &  68.6 & \gps & 22.72 & 0.15 &  PS1 \\
55680.3 &  70.7 & \gps & 22.94 & 0.17 &  PS1 \\
55716.8 &  96.7 & $g'$ & 23.35 & 0.06 &  GMOS \\
\hline
55545.5 & $-$25.3 & \rps & $>$24.04 & \nodata & PS1 \\
55560.6 & $-$14.5 & \rps & 23.39 & 0.32 &  PS1 \\
55563.6 & $-$12.4 & \rps & 22.97 & 0.25 &  PS1 \\
55566.6 & $-$10.3 & \rps & 22.01 & 0.07 &  PS1 \\
55572.6 &  $-$6.0 & \rps & 21.82 & 0.10 &  PS1 \\
55574.3 &  $-$4.8 & $r'$ & 21.55 & 0.03 &  LDSS3 \\
55587.6 &   4.7 & \rps & 21.53 & 0.09 &  PS1 \\
55590.5 &   6.8 & \rps & 21.62 & 0.05 &  PS1 \\
55593.5 &   8.9 & \rps & 21.77 & 0.05 &  PS1 \\
55596.4 &  11.0 & \rps & 21.62 & 0.06 &  PS1 \\
55602.4 &  15.2 & \rps & 21.95 & 0.22 &  PS1 \\
55627.1 &  32.8 & $r'$ & 22.60 & 0.07 &  IMACS \\
55629.3 &  34.4 & \rps & 22.48 & 0.09 &  PS1 \\
55632.3 &  36.5 & \rps & 22.71 & 0.14 &  PS1 \\
55635.3 &  38.7 & \rps & 22.71 & 0.11 &  PS1 \\
55650.4 &  49.4 & \rps & 22.68 & 0.13 &  PS1 \\
55653.4 &  51.5 & \rps & 22.94 & 0.21 &  PS1 \\
55662.3 &  57.9 & \rps & 22.66 & 0.15 &  PS1 \\
55674.3 &  66.4 & \rps & 22.83 & 0.10 &  PS1 \\
55677.3 &  68.6 & \rps & 23.17 & 0.18 &  PS1 \\
55680.3 &  70.7 & \rps & 23.13 & 0.28 &  PS1 \\
55716.8 &  96.7 & $r'$ & 23.15 & 0.10 &  GMOS \\
\hline
55546.5 & $-$24.6 & \ips & $>$23.47 & \nodata & PS1 \\
55561.6 & $-$13.8 & \ips & 22.88 & 0.19 &  PS1 \\
55567.5 &  $-$9.6 & \ips & 22.31 & 0.13 &  PS1 \\
55570.6 &  $-$7.4 & \ips & 21.87 & 0.10 &  PS1 \\
55576.6 &  $-$3.1 & \ips & 21.78 & 0.08 &  PS1 \\
55588.6 &   5.4 & \ips & 21.86 & 0.09 &  PS1 \\
55594.5 &   9.6 & \ips & 22.04 & 0.12 &  PS1 \\
55597.4 &  11.7 & \ips & 21.96 & 0.09 &  PS1 \\
55615.5 &  24.6 & \ips & $>$22.23 & \nodata & PS1 \\
55645.3 &  45.8 & \ips & 23.02 & 0.25 &  PS1 \\
55651.4 &  50.1 & \ips & 22.78 & 0.19 &  PS1 \\
55654.3 &  52.2 & \ips & 23.13 & 0.27 &  PS1 \\
55672.3 &  65.0 & \ips & 23.32 & 0.29 &  PS1 \\
55675.3 &  67.1 & \ips & 23.28 & 0.29 &  PS1 \\
55681.3 &  71.4 & \ips & 23.06 & 0.23 &  PS1 \\
55716.8 &  96.7 & $i'$ & $>$22.65 & \nodata & GMOS \\
\hline
55571.6 &  $-$6.7 & \zps & 21.95 & 0.30 &  PS1 \\
55577.6 &  $-$2.4 & \zps & 21.73 & 0.12 &  PS1 \\
55586.6 &   4.0 & \zps & 22.08 & 0.11 &  PS1 \\
55589.6 &   6.1 & \zps & 21.94 & 0.14 &  PS1 \\
55592.6 &   8.3 & \zps & 21.89 & 0.07 &  PS1 \\
55595.5 &  10.3 & \zps & 21.86 & 0.08 &  PS1 \\
55631.3 &  35.8 & \zps & 22.53 & 0.22 &  PS1 \\
55634.3 &  37.9 & \zps & 22.79 & 0.13 &  PS1 \\
55646.3 &  46.5 & \zps & 22.88 & 0.19 &  PS1 \\
55649.4 &  48.7 & \zps & $>$23.12 & \nodata & PS1 \\
55652.4 &  50.8 & \zps & $>$22.76 & \nodata & PS1 \\
55673.3 &  65.7 & \zps & 23.18 & 0.29 &  PS1 \\
55676.3 &  67.8 & \zps & $>$22.54 & \nodata & PS1 \\
55716.8 &  96.7 & $z'$ & $>$22.66 & \nodata & GMOS \\
\hline
55641.5 &  43.1 & \yps & 21.77 & 0.49 &  PS1 \\
55668.4 &  62.2 & \yps & $>$21.24 & \nodata & PS1 \\
55669.3 &  62.9 & \yps & $>$21.46 & \nodata & PS1 \\
55671.3 &  64.3 & \yps & $>$21.28 & \nodata & PS1 \\
\enddata
\tablenotetext{a}{In rest-frame days relative to maximum light on MJD
  55581.0. }
\tablenotetext{b}{Corrected for Galactic reddening.  Upper limits are 3$\sigma$.}
\label{phottab}
\end{deluxetable}

\begin{deluxetable*}{cccccccccc}
\tablecaption{Log of Spectroscopic Observations}
\tablehead{
\colhead{UT Midpoint} &
\colhead{Epoch\tablenotemark{a}} &
\colhead{Instrument} &
\colhead{Wavelength} &
\colhead{Slit} &
\colhead{Grating/} &
\colhead{Filter} &
\colhead{Exposure} &
\colhead{Mean} & 
\colhead{Position} \\
\colhead{(YYYY-MM-DD.DD)} &
\colhead{(days)} &
\colhead{} &
\colhead{Range (\AA)} &
\colhead{($\arcsec$)} &
\colhead{Grism} &
\colhead{} &
\colhead{Time (s)} &
\colhead{Airmass} &
\colhead{Angle (\degr)}
}
\startdata
2011-01-13.27 & $-5$ & LDSS3 & 3650$-$9460 & 0.75 & VPH-all & none &
1500 & 1.20 & 20 \\
2011-02-23.25 & 24 & Blue Channel & 3270$-$8500 & 1.0 & 300 & none &
5400 & 1.20 & 319 \\
2011-03-06.18 & 32 & IMACS & 4000$-$10100 & 0.9 & 300/$+$17.5 & none & 2400 &
1.20 & 167 \\
2011-03-07.17 & 33 & IMACS & 4000$-$10100 & 0.9 & 300/$+$17.5 & none & 3600 &
1.18 & 177 \\
2011-06-10.02 & 100 & GMOS-S & 3500$-$6280 & 1.0 & B600 & none & 3600
& 1.77 & 315 \\
\cutinhead{Host Spectra}
2011-12-29.39 & 244 & Blue Channel & 3320$-$8550 & 1.0 & 300 & none &
5400 & 1.22 & 135 \\
2013-01-11/14\tablenotemark{b} & 515 & LDSS3 & 3650$-$9450 & 0.75 & VPH-all & none &
2900 & 1.22 & 22/$-10$ \\
2013-01-12/13\tablenotemark{b} & 515 & LDSS3 & 5970$-$9300 & 1.0 & VPH-red & OG590 &
5700 & 1.20 & 17/$-10$ \\
\enddata
\tablenotetext{a}{In rest-frame days relative to maximum light.}
\tablenotetext{b}{Spectra taken on two nights.}
\label{spectab}
\end{deluxetable*} 

\begin{deluxetable}{lrcccc}
\tablecaption{\af\ VLA Observations}
\tablehead{
\colhead{MJD} &
\colhead{Epoch\tablenotemark{a}} &
\colhead{On-Source} & 
\colhead{Frequency} &
\colhead{3$\sigma$ Upper} \\
 & \colhead{(d)} & Time (min) & (GHz) & Limit ($\mu$Jy) 
}
\startdata
55649.0 & 48 & 36 & 4.9 & $<$51 \\
55933.4 & 251 & 17 & 5.5 & $<$30 \\
56444.0 & 614 & 60 & 5.5 & $<$45 \\
\enddata
\tablenotetext{a}{In rest-frame days relative to maximum light.}
\label{radtab}
\end{deluxetable}

\begin{deluxetable*}{cccccccc}
\tablecolumns{8}
\tablecaption{\af\ Host Galaxy Photometry}
\tablehead{
\colhead{Filter} &
\colhead{Observed} & \colhead {Rest-frame} &
\multicolumn{2}{c}{1\farcs15 Aperture} &
\multicolumn{2}{c}{3\arcsec Aperture} &
\colhead{Instrument} \\
 & \colhead{Wavelength (\AA)} & \colhead{Wavelength (\AA)} &
 AB mag\tablenotemark{a} & error & AB mag\tablenotemark{a} & error &
}
\startdata
FUV & 1539 & 1095 & \nodata & \nodata & $>$24.4\tablenotemark{b} & \nodata & \galex\ \\
NUV & 2316 & 1649 & \nodata & \nodata & $>$24.3 & \nodata & \galex\ \\
\gps & 4825 & 3424 & 23.42 & 0.04 & 22.68 & 0.04 & PS1 \\
\rps & 6170 & 4393 & 22.00 & 0.02 & 21.35 & 0.02 & PS1 \\
\ips & 7520 & 5354 & 21.48 & 0.02 & 20.87 & 0.02 & PS1 \\
\zps & 8660 & 6165 & 21.20 & 0.02 & 20.60 & 0.02 & PS1 \\
\yps & 9620 & 6849 & 21.07 & 0.03 & 20.45 & 0.04 & PS1 \\
$J$ & 12360 & 8798 & 20.64 & 0.03 & 20.02 & 0.03 & FourStar \\
$H$ & 16620 & 11830 & 20.38 & 0.02 & 19.86 & 0.02 & FourStar \\
\enddata
\tablenotetext{a}{Corrected for Galactic extinction.}
\tablenotetext{b}{Upper limits are 5$\sigma$.}
\label{hosttab}
\end{deluxetable*}

\end{document}